%% file: paper.tex
\documentclass[british,10pt,final,twocolumn,twoside]{IEEEtran}
\usepackage[T1]{fontenc}
\usepackage[utf8]{inputenc}
\usepackage{verbatim}
\usepackage{amsmath}
\usepackage{amssymb}
\usepackage{graphicx}

\makeatletter

\usepackage{acronym}
\usepackage{graphicx}
\usepackage{cite}       

\usepackage{booktabs}
\usepackage{multirow}

\usepackage{pgfplots}
\pgfplotsset{compat=newest}
\pgfplotsset{plot coordinates/math parser=false}
\usetikzlibrary{plotmarks}

\usepackage{exscale,relsize}

\global\long\def\step{\operatorname{u}}

\global\long\def\dbm{\,\text{dBm}}
\global\long\def\dbmm{\,\frac{\text{dBm}}{\text{m}}}
\global\long\def\dbi{\,\text{dBi}}
\global\long\def\mhz{\,\text{MHz}}
\global\long\def\ghz{\,\text{GHz}}
\global\long\def\meter{\,\text{m}}
\global\long\def\msq{\,\text{m}^2}
\global\long\def\bps{\,\text{bps}}

\global\long\def\toa{\text{\tiny{TOA}}}

\global\long\def\rss{\text{\tiny{RSS}}}

\global\long\def\los{\text{\tiny{LOS}}}
\global\long\def\nlos{\text{\tiny{NLOS}}}

\global\long\def\mymax{\text{\tiny{max}}}

\global\long\def\map{\text{\tiny{MAPBE}}}
\global\long\def\ml{\text{\tiny{MLE}}}

\newcommand{\executeiffilenewer}[3]{%
\ifnum\pdfstrcmp{\pdffilemoddate{#1}}%
{\pdffilemoddate{#2}}>0%
{\immediate\write18{#3}}\fi%
}

\newacro{GPS}{global positioning system}
\newacro{UWB}{ultra wide band}
\newacro{TOA}{time of arrival}
\newacro{TDOA}{time difference of arrival}
\newacro{AOA}{angle of arrival}
\newacro{RSS}{received signal strength}
\newacro{GIS}{geographical information systems}
\newacro{PF}{particle filter}
\newacro{BFI}{Bayesian Fisher information}
\newacro{BFIM}{Bayesian Fisher information matrix}
\newacro{CRB}{Cramer Rao bound}     
\newacro{BCRB}{Bayesian Cramer Rao bound}
\newacro{EZZB}{extended Ziv Zikai bound}
\newacro{WWB}{Weiss Weinstein bound}

\newacro{RMSE}{root median square error}                    

\newacro{MSE}{mean square error}
\newacro{BMSE}{Bayesian mean square error}

\newacro{MMSE}{minimum mean square error}
\newacro{MMSEE}{minimum mean square error estimator}

\newacro{MAP}{maximum a posteriori}
\newacro{ML}{maximum likelihood}
\newacro{MAPBE}{maximum a posteriori Bayesian estimator}
\newacro{MLE}{maximum likelihood estimator}

\newacro{SNR}{signal-to-noise ratio}
\newacro{FLOP}{floating point operation}
\newacro{NLOS}{non-line-of-sight}
\newacro{HMM}{hidden Markov model}
\newacro{KF}{Kalman filter}
\newacro{PCRB}{posterior Cramer Rao bound}
\newacro{PFI}{posterior Fisher information}
\newacro{PFIM}{posterior Fisher information matrix}
\newacro{SLAM}{simultaneous localization and mapping}
\newacro{LS}{least-squares}
\newacro{RHS}{right hand side}

\newacro{POCS}{projection onto convex sets}
\newacro{MDS}{multidimensional scaling}

\newacro{GDOP}{geometric dilution of precision}

\newacro{LOS}{line of sight}
\newacro{WED}{wall extra delay}
\newacro{MNC}{minimal non-overlapping covering}
\newacro{DNC}{``dense'' non-overlapping covering}
\newacro{RRS}{reproducible research standard}
\newacro{DRD-MMSE}{distance-reduced domain MMSE}
\newacro{DRD-MAP}{distance-reduced domain MAP}
\newacro{PRD-MMSE}{probability-reduced domain MMSE}
\newacro{PRD-MAP}{probability-reduced domain MAP}
\newacro{RDE}{room detection error}

\newacro{AF}{attenuation factor}
\newacro{WAF}{wall attenuation factor}
\newacro{MWM}{multi-wall model}

\newlength\figureheight
\newlength\figurewidth

\makeatother

\usepackage[british]{babel}

\begin{document}

\title{Map-Aware Models for Indoor Wireless Localization Systems: An Experimental Study}

\author{Francesco Montorsi,~\IEEEmembership{Student Member,~IEEE,} Fabrizio
Pancaldi,~\IEEEmembership{Member,~IEEE}, \\
Giorgio M. Vitetta,~\IEEEmembership{Senior Member,~IEEE}%
\thanks{Francesco Montorsi and Giorgio M. Vitetta are with the Dept. of Engineering
``Enzo Ferrari'', University of Modena and Reggio Emilia, Modena,
Italy (e-mail: francesco.montorsi@unimore.it; giorgio.vitetta@unimore.it).%
}%
\thanks{Fabrizio Pancaldi is with the Dept. of Science and Methods for Engineering,
University of Modena and Reggio Emilia, Reggio Emilia, Italy (e-mail:
fabrizio.pancaldi@unimore.it).%
}}
\maketitle
\begin{abstract}
The accuracy of indoor wireless localization systems can be substantially
enhanced by \emph{map-awareness}, i.e., by the knowledge of the map
of the environment in which localization signals are acquired. In
fact, this knowledge can be exploited to cancel out, at least to some
extent, the signal degradation due to propagation through physical
obstructions, i.e., to the so called \emph{non-line-of-sight} bias.
This result can be achieved by developing novel localization techniques
that rely on proper map-aware statistical modelling of the measurements
they process. In this manuscript a unified statistical model for the
measurements acquired in map-aware localization systems based on \emph{time-of-arrival
}and \emph{received signal strength} techniques is developed and its
experimental validation is illustrated. Finally, the accuracy of the
proposed map-aware model is assessed and compared with that offered
by its map-unaware counterparts. Our numerical results show that,
when the quality of acquired measurements is poor, map-aware modelling
can enhance localization accuracy by up to 110\% in certain scenarios.\end{abstract}

\section{Introduction\label{sec:intro}}

In recent years significant attention has been devoted to the development
of accurate and low cost wireless localization systems for indoor
civilian applications, since they can be employed to provide a number
of new services, like asset tracking and tracking of people with special
needs \cite{Patwari2005,Pahlavan2002}. In these services estimated
positions need always to be related to a surrounding infrastructure
(e.g., rooms and corridors) to be useful to their end users, so that
the knowledge of the map of the environment where users are expected
to move (e.g., the plan of a building floor) is required. In principle,
map knowledge (i.e., \emph{map-awareness}) can be also exploited to
improve the estimation accuracy of a localization system. In fact,
any wireless localization system first acquires a set of point-to-point
measurements related to user position (first step; technology-dependent)
and then processes such measurements for \emph{bi-dimensional} (2-D)
or three-dimensional position estimation by means of a proper localization
technique (second step; technology-agnostic) \cite[Sec. 4]{drb1}.
Maps can play a significant role in the second step, since they provide
information about environmental obstructions (e.g., walls) which interfere
with signal propagation; however, a full exploitation of these information
requires a) the availability of \emph{map-aware statistical models}
for the acquired measurements and b) the development of localization
techniques explicitly based on these models. 

At present the only available map-aware statistical models are the
\ac{WED} model \cite[eq. (6)]{Dardari2008a} and the \ac{AF}
model \cite[Sec. 4.11.5]{Rappaport2002} (also known as \emph{wall-attenuation}
model or \emph{multi-wall} model \cite{Lott2001}); these models have
been developed for \ac{TOA} and \ac{RSS} localization systems,
respectively, and are based on experimental evidence. This preliminary
work shows that map-awareness can significantly improve localization
accuracy by compensating for the so called \ac{NLOS} \emph{bias},
which is a major source of error. However, as far as we know, the
accuracy and validity of the above mentioned models in real world
localization systems is under-explored and, generally speaking, there
is a lack of experimental results supporting them in the technical
literature. In fact, most of the state-of-the-art localization methods
rely on \textit{map-unaware models}. For instance, the well known
\textit{log-distance propagation model} \cite{Patwari2005,Li2007a,Yu2009,Venkatesh2007a,Mazuelas2009b,Rappaport2002}
(or, in some cases, models based on polynomial series expansions \cite{Yang2009})
are adopted to relate RSS to distance. Similar comments hold for those
models relating \ac{TOA} and \ac{TDOA} to distance; in this
case additive error terms are usually represented by Gaussian \emph{random
variables} (rvs) \cite[eq. (6)]{Patwari2005}, \cite{Alavi2006,Venkatesh2007,Jourdan2006},
although more refined models accounting for \ac{NLOS} propagation
are also available (see \cite{Montorsi_NLOS_ADVET_JOURNAL,Montorsi_PHD_THESIS}
and references therein).

It is also worth mentioning that map-awareness is implicitly employed
in \emph{fingerprinting}-based localization systems to select fingerprint
locations \cite{Fang2008,Guerrero-curieses2009,Bshara2010}. In those
systems no statistical modelling of acquired measurements is needed
(even if combined fingerprinting/statistical approaches are possible
\cite{Wang2011}); however, extensive and time consuming measurement
campaigns (which may be very sensitive to environmental changes) are
necessary to achieve an acceptable accuracy, since the localization
error of fingerprinting methods is roughly bounded by the spacing
between calibration sites.

The aim of this manuscript is twofold. First of all, a novel\emph{
unified statistical map-aware model} for \ac{TOA}, \ac{TDOA}
or \ac{RSS} measurements is proposed and is validated exploiting
a set of RSS and \ac{UWB} TOA data acquired in indoor environments.
Secondly, the improvement in localization accuracy provided by optimal
localization algorithms based on the novel map-aware modelling with
respect to their counterparts relying on map-unaware modelling is
quantified; this unveils that, specially in RSS systems, the accuracy
improvement justifies the increased complexity of map-aware modelling. 

The proposed map-aware model has the following relevant features:
a) it relates the NLOS bias affecting measurements to map geometrical
features; b) it can be employed in localization systems based on ranging
techniques (i.e., TOA, TDOA and RSS, but not angle of arrival \cite{Liu2007,drb1})
provided that their radio signals mainly propagate through obstructions;
c) it contains few parameters to be estimated from measurements; d)
even if its validity is assessed for narrowband low-frequency RSS
measurements and for TOA \ac{UWB} measurements, its use can be
envisaged for other technologies, like \emph{wireless local area network}
TOA/TDOA or even non-radio-based technologies (e.g, ultra-sound),
since it does not rely on technology-specific properties; e) likelihood
functions based on it can be employed in navigation systems (where
mobile agents are considered). 

This manuscript is organized as follows. In Section \ref{sec:statistical_model},
the localization system we consider is described and general statistical
models for map-aware and map-unaware scenarios are proposed. Specific
models for RSS and TOA measurements, based on our experimental data,
are illustrated in Section \ref{sec:experimental_results}. In Section
\ref{sec:accuracy_results} map-aware and map-unaware optimal localization
algorithms are derived, their performance is assessed and compared,
and some indications about their computational complexity are provided.
Finally, Section \ref{sec:conclusions} offers some conclusions.

\emph{Notations}: The probability density function (pdf) of the rv
$R$ evaluated at the point $r$ is denoted $f(r)$; $\mathcal{N}\left(r;\mu,\sigma^{2}\right)$
denotes the pdf of a Gaussian rv having mean $\mu$ and variance $\sigma^{2}$,
evaluated at the point $r$; $\mathcal{E}\left(r;t,\mu\right)$ denotes
the pdf of an exponential rv translated by $t$ and having mean $t+\mu$,
evaluated at the point $r$, so that $\mathcal{E}\left(r;t,\mu\right)\triangleq\frac{1}{\mu}\exp\left(\frac{t-r}{\mu}\right)\step(r-t)$,
where $\step(x)$ denotes the Heaviside step function; $\left|\mathcal{S}\right|$
denotes the cardinality of the set $\mathcal{S}$ ;$\left\lfloor x\right\rfloor $
denotes the floor of the real parameter $x$; $\star$ denotes the
convolution integral.

\section{Modelling of Reference Scenario\label{sec:statistical_model}}

\begin{figure}
\centering
\def\svgwidth{6cm}
\executeiffilenewer{fig1.svg}{fig1.pdf}%
{inkscape  -z  -D  --file=fig1.svg  %
--export-pdf=fig1.pdf  --export-latex}%
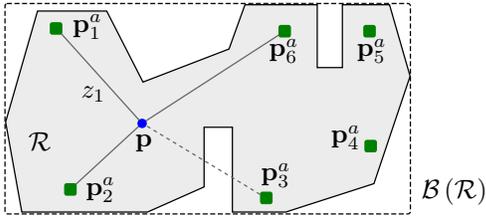%

\caption{Example of a localization system employing $N_{a}=6$ anchors (green
squares) to localise a single agent (blue circle) and operating over
a bounded map (whose support $\mathcal{R}$ is identified by the grey
area). The presence of a LOS (NLOS) link between the agent and a given
anchor is evidenced by a continuous (dashed) line (note that propagation
conditions prevent establishing a wireless link between the agent
and the anchors 4 and 5).\label{fig:system_model}}
\end{figure}
In the following we focus on a 2-D localization system operating in
an indoor scenario and employing $N_{a}$ devices, called \emph{anchors}
and whose positions $\left\{ \mathbf{p}_{i}^{a}\right\} _{i=1}^{N_{a}}$
are known, to estimate the position $\mathbf{p}$ of a single \emph{agent}
in a large indoor scenario (i.e., the floor of a large building).
The system structure is exemplified by Fig. \ref{fig:system_model}
for $N_{a}=6$. Formulating the problem of \emph{map-aware localization}
requires defining mathematical models for a) its map, b) the wireless
connectivity between the agent and the anchors, and c) the measurements
acquired for position estimation. These models are illustrated in
the following three paragraphs.

\subsection{Map Model\label{sub:map_model}}

Our prior knowledge about the agent position $\mathbf{p}\triangleq\left[x,y\right]^{T}$
is expressed by the \emph{uniform} pdf \cite{Montorsi_APRIORI_BOUNDS_TIT}
\begin{equation}
f(\mathbf{p})=\begin{cases}
1/\mathcal{A_{R}} & \mathbf{p}\in\mathcal{R}\\
0 & \text{elsewhere}
\end{cases}\label{eq:map_pdf}
\end{equation}
where $\mathcal{R}\subset\mathbb{R}^{2}$ denotes the region where
the agent is constrained to lie and $\mathcal{A}_{\mathcal{R}}$ its
area. Note that (\ref{eq:map_pdf}) describes a \emph{uniform map}
model, which is fully characterized by the map \emph{support }$\mathcal{R}$
(whose \emph{bounding box }is $\mathcal{B}\left(\mathcal{R}\right)$).
In the following we assume that any \emph{map-aware} (\emph{map-unaware})
localization system is endowed with the knowledge of $\mathcal{R}$
($\mathcal{B}\left(\mathcal{R}\right)$). Note that the propagation
of wireless signals (and, consequently, localization performance)
is affected by the obstructions (e.g., walls) shaping the map support
$\mathcal{R}$; in particular, an important role is played by the
number of obstructions $N_{o}\left(\mathbf{p}_{1},\mathbf{p}_{2}\right)$
interposed between two arbitrary points $\mathbf{p}_{1}$ and $\mathbf{p}_{2}$,
both belonging to $\mathcal{R}$.

\subsection{Connectivity Model\label{sub:connectivity_model}}

We assume that an agent located at $\mathbf{p}$ is connected with
the $i$-th anchor when the \ac{SNR} of the wireless signal radiated
by this anchor and received by the agent exceeds a given threshold
(i.e., the agent's \emph{receiver sensitivity}); therefore, each anchor
is characterized by a specific coverage region. Even if the map of
the environment is known, the prediction of the shape of each coverage
region is not easy. For this reason, the following simple (map-unaware)
connectivity model is adopted: the coverage region $\mathcal{R}^{(i)}$
of the $i$-th anchor is a circle centered at $\mathbf{p}_{i}^{a}$
and whose radius $d_{\mymax,i}$ depends on the power radiated by
the anchor itself and the propagation conditions%
\footnote{The availability of $d_{\mymax,i}$ represents a form of a priori
knowledge; when such a knowledge is unavailable, $d_{\mymax,i}=+\infty$
can be selected.%
}. Then, the signal coming from the $i$-th anchor is \emph{theoretically}
detectable inside $\mathcal{R}^{(i)}$ and in such a region some localization
information or, more precisely, an \emph{observation} (denoted $z_{i}$)
can be acquired in the first step of any localization technique (see
Section \ref{sec:intro}). 

In practice, in the presence of harsh propagation conditions the number
$N_{obs}$ of observations available to the agent may differ from
the number which can be predicted theoretically using the above mentioned
connectivity model. Let $\mathcal{Z}$ denote the set of indices associated
with the anchors \emph{truly connected} with the agent; then, we have
that $N_{obs}\triangleq\left|\mathcal{Z}\right|$ and that, if $i\in\mathcal{Z}$,
then $\mathbf{p}\in\mathcal{R}^{(i)}\bigcap\mathcal{R}$. Consequently,
in a \emph{map-aware} localization system the agent position has to
be searched for inside the region 
\begin{equation}
\mathcal{R}^{(\mathcal{Z})}\triangleq\left(\bigcap_{i\in\mathcal{Z}}\mathcal{R}^{(i)}\right)\bigcap\mathcal{R}\label{eq:mapaware_anchor_coverage_region}
\end{equation}
whereas, in a \emph{map-unaware} system, the agent position is expected
to belong to 
\begin{equation}
\mathcal{P}^{(\mathcal{Z})}\triangleq\left(\bigcap_{i\in\mathcal{Z}}\mathcal{R}^{(i)}\right)\bigcap\mathcal{B}\left(\mathcal{R}\right)\label{eq:nomap_anchor_coverage_region}
\end{equation}
Finally, it is important to note that the wireless links referring
to the anchors connected with an agent located at $\mathbf{p}$ can
be in \ac{LOS} or \ac{NLOS} conditions. For this reason, the
set $\mathcal{Z}$ can be partitioned into the subsets $\mathcal{Z}^{\los}\triangleq\left\{ i\in\mathcal{Z}|N_{o}\left(\mathbf{p},\mathbf{p}_{i}^{a}\right)=0\right\} $
and $\mathcal{Z}^{\nlos}\triangleq\mathcal{Z}/\mathcal{Z}^{\los}$
containing the indices of the connected anchors in \ac{LOS} and
\ac{NLOS} conditions, respectively.

\subsection{Observation Model\label{sub:observation_model}}

The localization system we consider statistically infers the agent
position $\mathbf{p}$ from a set of $N_{obs}$ observations (in practice,
estimated ranges) $\left\{ z_{i},i\in\mathcal{Z}\right\} $ extracted
from the radio signals coming from the connected anchors. For an indoor
environment, such signals are affected by \emph{multipath fading}
\cite[Sec. 5]{Rappaport2002} which substantially complicates the
problem of the statistical modelling of the available observations.
To simplify this problem, we assume that: a) the time interval needed
to acquire all the observations is small with respect to the coherence
time of the propagation scenario, so that the effects due to its \emph{time
selectivity} can be ignored; b) in TOA-based systems the effects due
to the \emph{time dispersion} characterising the propagation scenario
are mitigated transmitting wideband signals, so that multiple echoes
of the signal radiated by a given anchor can be resolved by the agent
(in particular, the first path can be correctly detected). Given such
assumptions, the observation model 
\begin{equation}
z_{i}=d_{i}(\mathbf{p})+b_{i}(\mathbf{p})+n_{i}(\mathbf{p})\label{eq:z_model}
\end{equation}
characterized by additive bias and noise terms and commonly adopted
in the localization literature (e.g., see \cite[Eq. (3)]{Dardari2008a},
\cite{Venkatesh2007a} and \cite[Eq. (3)]{Venkatesh2007} ) can be
also employed in our scenario for any $i\in\mathcal{Z}$; here $d_{i}(\mathbf{p})\triangleq\left\Vert \mathbf{p}-\mathbf{p}_{i}^{a}\right\Vert $
denotes the \emph{true distance} between the agent position $\mathbf{p}$
and the $i$-th anchor, $b_{i}(\mathbf{p})$ represents the NLOS \emph{bias}
originating from obstructions affecting the $i$-th wireless link
and $n_{i}(\mathbf{p})\sim\mathcal{N}(0,\sigma_{n,i}^{2}(\mathbf{p}))$
is a noise term modelling various sources of errors (e.g., the error
introduced in extrapolating the observation $z_{i}$ from the continuous-time
waveform collected by the agent antenna, the radio interference, and
the noise generated by both transmitter and receiver hardware). It
is important to point out that:
\begin{enumerate}
\item The effects of \emph{spatial selectivity }(that may generate substantial
fluctuations in the received power as the agent position changes)
are difficult to predict and hence have been included in the noise
term \foreignlanguage{english}{$n_{i}(\mathbf{p})$}, even if the
availability of a simple and reliable model for spatial fading would
certainly improve model accuracy.
\item The $i$-th observation $z_{i}$ (\ref{eq:z_model}) is generated
by mapping, through a function denoted $\psi(\cdot)$ in the following,
the physical quantity measured by the hardware (e.g., a RSS or a TOA)
into a \emph{range}. The function $\psi(\cdot)$ is derived under
the assumption of LOS propagation, since NLOS propagation conditions
are accounted for by introducing the bias term $b_{i}(\mathbf{p})$
in (\ref{eq:z_model}). In map-aware modelling such a term is modelled
as $b_{i}(\mathbf{p})\sim\mathcal{N}(\mu_{b,i}(\mathbf{p}),\sigma_{b,i}^{2}(\mathbf{p}))$
for $i\in\mathcal{Z}^{\nlos}$ and $b_{i}(\mathbf{p})=0$ for $i\in\mathcal{Z}^{\los}$;
a different (and not necessarily Gaussian) model is adopted, instead,
in map-unaware modelling, as discussed in detail below. 
\end{enumerate}
Providing further details about the observation model (\ref{eq:z_model})
requires a) considering specific localization techniques (in the following,
quantities specifically referring to TOA and RSS techniques are identified
by the superscripts $^{\toa}$ and $^{\rss}$, respectively) and b)
making a clear distinction between map-aware and map-unaware modelling,
as illustrated below. 

TOA\emph{-based localization} - In this case the observation $z_{i}$
is related to the measured TOA \foreignlanguage{english}{$\tau_{i}$}
(expressed in seconds) by the LOS mapping $z_{i}^{\toa}=\psi^{\toa}(\tau_{i})=c_{0}\tau_{i}$,
where $c_{0}$ is the speed of light in vacuum. In NLOS propagation
conditions $z_{i}^{\toa}$ is affected by a \emph{bias} (due to the
fact that obstacles slow down electromagnetic waves); in map-aware\emph{
}systems, the NLOS bias mean $\mu_{b,i}^{\toa}(\mathbf{p})$ can be
modelled as \cite[Sec. 3.3.1]{Dardari2008a}
\begin{equation}
\mu_{b,i}^{\toa}(\mathbf{p})=\sum_{j=1}^{N_{o}\left(\mathbf{p},\mathbf{p}_{i}^{a}\right)}t_{w,j}\sqrt{\epsilon_{r,w,j}-1}\label{eq:bias_mean_TOA}
\end{equation}
for any $i\in\mathcal{Z}$, where $t_{w,j}$ and $\epsilon_{r,w,j}$
are the thickness and the relative electrical permittivity, respectively,
of the $j$-th obstruction. If all the obstructions have the same
thickness $t_{w}$ and the same permittivity $\epsilon_{r,w}$, (\ref{eq:bias_mean_TOA})
simplifies as $\mu_{b,i}^{\toa}(\mathbf{p})=N_{o}\left(\mathbf{p},\mathbf{p}_{i}^{a}\right)\cdot t_{w}\sqrt{\epsilon_{r,w}-1}$
\cite[Eq. (6)]{Dardari2008a}. In map-unaware localization systems,
instead, we assume that: a) a state-of-the-art localization strategy
relying on a LOS/NLOS detector is adopted%
\footnote{LOS/NLOS detectors consist in algorithms estimating the LOS or NLOS
condition based on the received signal only; they often work analysing
the evolution in time of the received signal; see \cite{Montorsi_NLOS_ADVET_JOURNAL}
and references therein for more details.%
} ; b) such a detector is characterized by an error (i.e., false detection
and missed detection) probability $P_{e}^{\nlos}$ over each link;
c) it provides the map-unaware localization algorithm with the estimates
$\hat{\mathcal{Z}}^{\los}$ and $\hat{\mathcal{Z}}^{\nlos}$ of the
sets $\mathcal{Z}^{\los}$ and $\mathcal{Z}^{\nlos}$, respectively.
Given these estimates, the model $b_{i}^{\toa}(\mathbf{p})\sim\mathcal{E}(0,\nu_{b}^{\toa})$
is employed for any $i\in\hat{\mathcal{Z}}^{\nlos}$, since the NLOS
bias is always positive for TOA measurements\textbf{ }\cite{Venkatesh2007,Jourdan2006,Guvenc2009}.

TDOA\emph{-based localization} - The results illustrated above for
TOA modelling can be easily exploited for TDOA-based localization
systems. In fact, in this case $z_{i}$ is evaluated subtracting the
TOA provided by a \emph{reference} anchor from that acquired by the
$i$-th anchor and both measurements are modelled by (\ref{eq:z_model});
further details are not provided here for space limitations.

RSS\emph{-based localization} - In this case the observation $z_{i}$
is related to the measured RSS $P_{rx,i}$ (i.e., the average power
of the radio signal received by the agent from the $i$-th anchor)
according to the formula $z_{i}^{\rss}=\psi^{\rss}(P_{rx,i})$; specific
forms for $\psi^{\rss}(\cdot)$ are discussed later in Sec. \ref{sub:rss_measurements}.
NLOS propagation needs to be accounted for carefully in RSS-based
systems, since the attenuation due to obstructions results in an overestimated
range $z_{i}^{\rss}$. However, unlike the TOA and TDOA cases, no
simple and accurate theoretical formula is available to model the
NLOS bias mean in map-aware localization, so that models extracted
from experimental data must be employed. In our work measurement-based
models have been developed for both $\mu_{b,i}^{\rss}(\mathbf{p})$
and $\sigma_{b,i}^{\rss}(\mathbf{p})$ (see Section \ref{sec:experimental_results});
such models are similar to the experimental \ac{AF} model (developed
in \cite[Sec. 4.11.5]{Rappaport2002} and assuming a logarithmic dependence
of the RSS on the number of obstructions). In the map-unaware case,
instead, we assume again the availability of a LOS/NLOS detector characterized
by the error probability $P_{e}^{\nlos}$ and adopt\textbf{ }the bias
model $b_{i}^{\rss}(\mathbf{p})\sim\mathcal{N}(\kappa_{b}^{\rss},\gamma_{b}^{\rss})$.

The statistical models illustrated above allow us to develop a novel
\emph{unified} (i.e., valid for TOA, TDOA and RSS) statistical representation
for the set of observations $\left\{ z_{i},i\in\mathcal{Z}\right\} $.
In fact, given the trial agent position $\tilde{\mathbf{p}}\in\mathcal{R}^{(\mathcal{Z})}$,
the \emph{map-aware} likelihood associated with these observations
can be expressed as (see (\ref{eq:z_model}))
\begin{equation}
f(\mathbf{z}|\tilde{\mathbf{p}})=\prod_{i\in\mathcal{Z}}\mathcal{N}\left(z_{i};d_{i}(\tilde{\mathbf{p}})+\mu_{b,i}(\tilde{\mathbf{p}}),\sigma_{n,i}^{2}(\tilde{\mathbf{p}})+\sigma_{b,i}^{2}(\tilde{\mathbf{p}})\right)\label{eq:likelihood_with_maps}
\end{equation}
where $\mathbf{z}\triangleq\left[z_{i},\forall i\in\mathcal{Z}\right]^{T}$.
Similarly, the \emph{map-unaware} counterpart of (\ref{eq:likelihood_with_maps}),
given the trial agent position $\tilde{\mathbf{p}}\in\mathcal{P}^{(\mathcal{Z})}$,
can be put in the form
\begin{align}
f(\mathbf{z};\tilde{\mathbf{p}})= & \prod_{i\in\hat{\mathcal{Z}}^{\los}}\mathcal{N}\left(z_{i};d_{i}(\tilde{\mathbf{p}}),\sigma_{n,i}^{2}(\tilde{\mathbf{p}})\right)\nonumber \\
 & \prod_{i\in\hat{\mathcal{Z}}^{\nlos}}\mathcal{N}\left(z_{i};d_{i}(\tilde{\mathbf{p}})+\kappa_{b},\sigma_{n,i}^{2}(\tilde{\mathbf{p}})+\gamma_{b}^{2}\right)\label{eq:likelihood_no_maps_gaussian}
\end{align}
or
\begin{align}
f(\mathbf{z};\tilde{\mathbf{p}})= & \prod_{i\in\hat{\mathcal{Z}}^{\los}}\mathcal{N}\left(z_{i};d_{i}(\tilde{\mathbf{p}}),\sigma_{n,i}^{2}(\tilde{\mathbf{p}})\right)\nonumber \\
 & \prod_{i\in\hat{\mathcal{Z}}^{\nlos}}\left[\mathcal{E}\left(z_{i};0,\nu_{b}\right)\star\mathcal{N}\left(z_{i};d_{i}(\tilde{\mathbf{p}}),\sigma_{n,i}^{2}(\tilde{\mathbf{p}})\right)\right]\label{eq:likelihood_no_maps_exp_exact}
\end{align}
if a Gaussian or an exponential model is adopted for the NLOS bias,
respectively. In the latter case, the convolution integral appearing
in the right-hand side of (\ref{eq:likelihood_no_maps_exp_exact})
substantially complicates the likelihood (\ref{eq:likelihood_no_maps_exp_exact});
for this reason in the technical literature (e.g., see \cite{Guvenc2009})
it is usually assumed that the NLOS bias dominates over the Gaussian
noise, so that (\ref{eq:likelihood_no_maps_exp_exact}) can be approximated
as 
\begin{align}
f(\mathbf{z};\tilde{\mathbf{p}})\simeq & \prod_{i\in\hat{\mathcal{Z}}^{\los}}\mathcal{N}\left(z_{i};d_{i}(\tilde{\mathbf{p}}),\sigma_{n,i}^{2}(\tilde{\mathbf{p}})\right)\prod_{i\in\hat{\mathcal{Z}}^{\nlos}}\mathcal{E}\left(z_{i};d_{i}(\tilde{\mathbf{p}}),\nu_{b}\right).\label{eq:likelihood_no_maps_exp}
\end{align}
Finally, it is important to stress that (\ref{eq:likelihood_with_maps})
and (\ref{eq:likelihood_no_maps_gaussian})-(\ref{eq:likelihood_no_maps_exp})
represent \emph{general} statistical models applying to TOA, TDOA
and RSS measurements (under the assumptions described above); in the
next Section they will be \emph{specialized }to fit our experimental
results.

\section{Extraction of Model Parameters From Experimental Results\label{sec:experimental_results}}

In this Section we provide a) various details about the measurement
campaigns supporting our proposed models, b) some explicit expressions
for the functions $\psi^{\rss}(\cdot)$, $\mu_{b,i}^{\rss}(\mathbf{p})$,
$\sigma_{b,i}^{2}(\mathbf{p})$ and $\sigma_{n,i}^{2}(\mathbf{p})$,
and c) the exact form of the likelihood functions fitting our experimental
data.

\subsection{RSS Measurement Campaign\label{sub:rss_measurements}}

RSS data have been acquired in a measurement campaign performed by
our research group in the first half of 2012. Two different types
of radio devices operating in two distinct frequency bands (namely,
$169\mhz$ and $2.4\ghz$) have been employed. In this paragraph a
detailed analysis of the results extracted from the data acquired
at $169\mhz$ only is provided; we comment on the differences between
theses results and those referring to $2.4\ghz$ in the following
paragraph. 

In our measurement campaign two EMB-WMB169T narrowband radio modules
based on Texas Instruments transceivers have been employed \cite{embit_169mhz_radios_datasheet}.
These modules have the following relevant features: a) they transmit
at $4800\bps$ employing a 2-GFSK modulation (whose bandwidth is about
3 kHz); b) they radiate $+15\dbm$ at the carrier frequency $f_{c}=169\mhz$;
c) their antenna is a dipole omnidirectional in the azimuthal plane
and $0.1\meter$ long (it is shorter than the wavelength $\lambda_{169\mhz}=1.77\meter$
for practical reasons) and its gain is equal to $-1\dbi$. 

\begin{figure}
\begin{centering}
\centering
\def\svgwidth{7cm}
\executeiffilenewer{fig2.svg}{fig2.pdf}%
{inkscape  -z  -D  --file=fig2.svg  %
--export-pdf=fig2.pdf  --export-latex}%
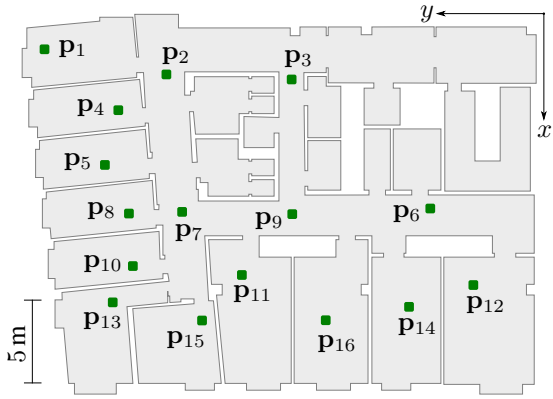%

\par\end{centering}

\centering{}\caption{Map support $\mathcal{R}$ of the floor where RSS measurements have
been acquired. The $N_{pos}^{\rss}$ measurement sites are indicated
by green squares.\label{fig:map_dii}}
\end{figure}

\begin{figure}
\centering{}\includegraphics[width=6cm]{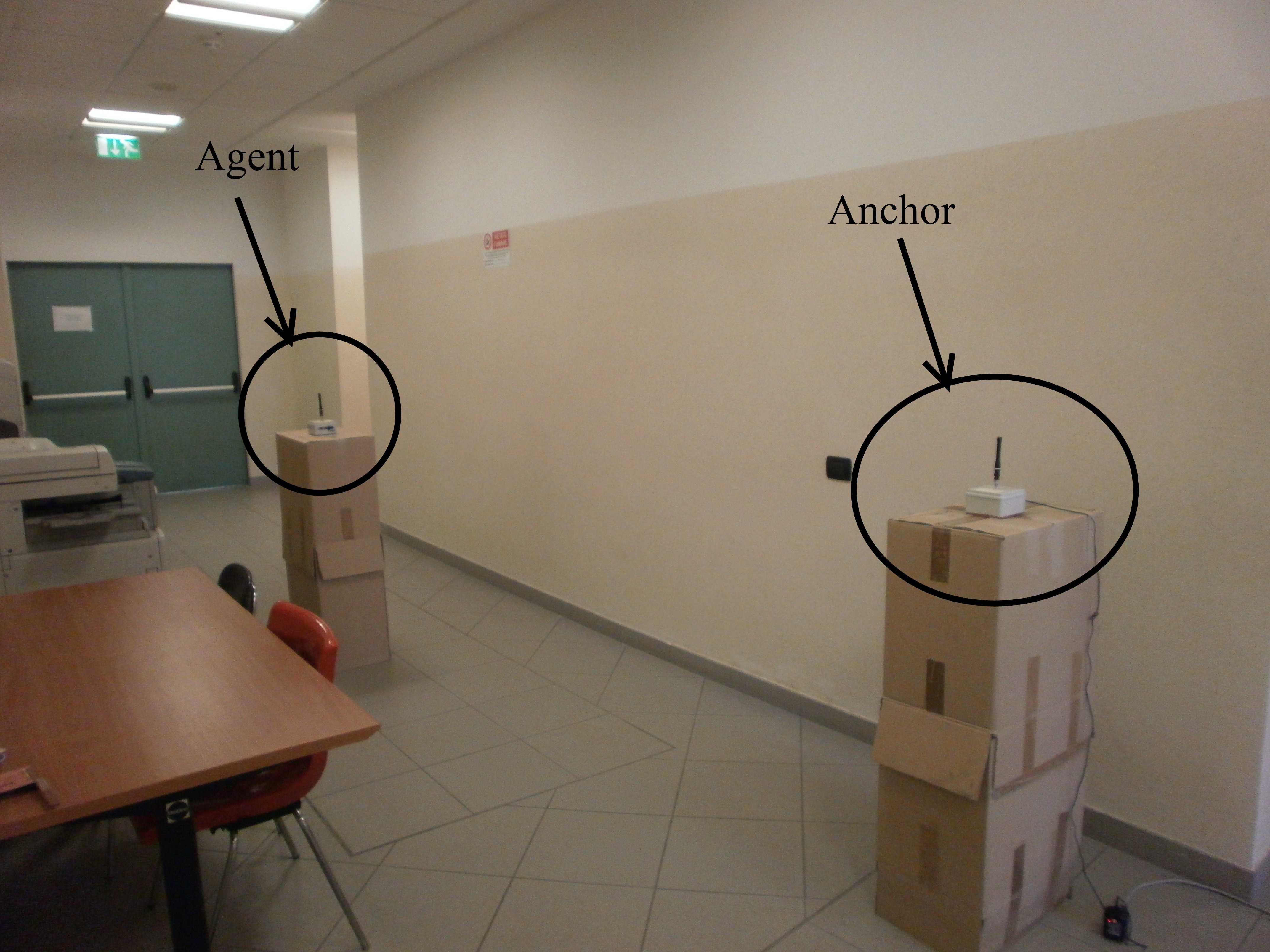}\caption{Adopted experimental setup for a LOS link; the employed radio devices
are located on the top of two paper boxes.\label{fig:expsetup_picture}}
\end{figure}

Various measurements have been acquired in both LOS and NLOS conditions.
The procedure we adopted was similar to that described in \cite{wprb_database}.
On the 2nd floor of our departmental building (see Fig. \ref{fig:map_dii};
the involved area was about $600\msq$) $N_{pos}^{\rss}=16$ different
``measurement sites'' have been selected; let $\mathbf{p}_{m}^{\rss}$
denote the position of the $m$-th site, with $m=1,...,N_{pos}^{\rss}$.
Then, the above mentioned radio modules were employed to acquire RSS
measurements referring to all $\binom{16}{2}=120$ location pairs.
In four of these pairs no wireless connection was established, yielding
116 measured links (the number%
\footnote{The value of $N_{obs}^{\rss}(m)$ depends on the coverage region of
each anchor, which, in turn, is affected by both a) the power radiated
by the anchor itself and b) the propagation environment (see our connectivity
model of paragraph \ref{sub:connectivity_model}).%
} of measurement sites ``visible'' from the $m$-th site will be
denoted as $N_{obs}^{\rss}(m)$) For each link (i.e., couple of sites),
$N_{z}=50$ separate RSS measurements were acquired (our experimental
setup is shown in Fig. \ref{fig:expsetup_picture}). Such measurements
have been stored in a database%
\footnote{All these acquired data and the MATLAB code developed to process them
are publicly available at \texttt{\scriptsize{http://frm.users.sf.net/publications.html}},
in accordance with the philosophy of reproducible research standard
\cite{Vandewalle2009}.\texttt{\footnotesize{ }}%
} and processed to estimate the functions $\mu_{b,i}^{\rss}(\mathbf{p})$,
$\sigma_{b,i}^{2}(\mathbf{p})$ and $\sigma_{n,i}^{2}(\mathbf{p})$,
as illustrated in detail in Appendix \ref{app:mapaware_fitting}.
Note that the generation of such a database required a density of
measurements (1 measurement site every $\sim40\msq$) much lower than
that required by fingerprinting methods (e.g., in \cite{Alippi2005}
1 measurement site every $\sim0.9\msq$ was necessary). 
 
\begin{figure}
\centering
\setlength\figureheight{5cm}
\setlength\figurewidth{7cm}
\input{fig4.tikz}

\caption{RSS versus distance in LOS conditions: experimental data and two different
regression fits are shown. The error bars indicate the standard deviation
of the measurements associated with each data point.\label{fig:rss_fitting_169mhz}}
\end{figure}
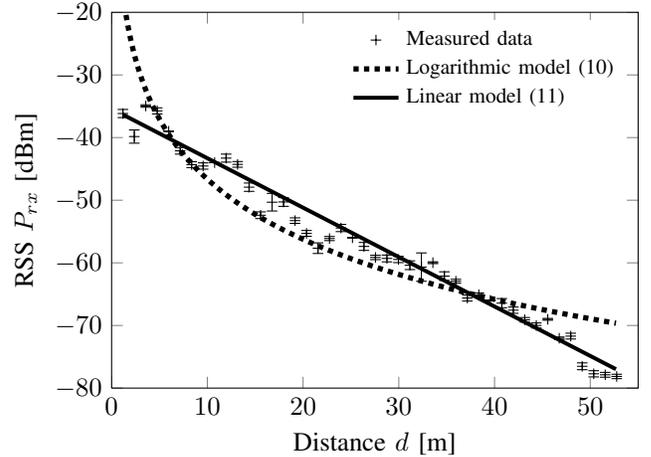

Additional RSS measurements have been acquired in LOS conditions to
estimate the function $\psi^{\rss}\left(P_{rx,i}\right)$ (see paragraph
\ref{sub:observation_model}). In particular, about 2200 RSS measurements
have been acquired in a long corridor of the same environment for
44 distinct values of $z^{\rss}$. Fig. \ref{fig:rss_fitting_169mhz}
shows such experimental results and their \ac{LS} regression fits
referring to a) the log-distance path loss model (see \cite{Patwari2005,Li2007a,Yu2009,Venkatesh2007a,Mazuelas2009b,Rappaport2002})
\begin{equation}
P_{rx}\left(z^{\rss}\right)=P_{0}-10\eta\log_{10}\frac{z^{\rss}}{d_{0}}\label{eq:logdistance_path_loss}
\end{equation}
where $P_{0}$ denotes the RSS measured at a reference\emph{ }distance
$d_{0}$ from the transmitter and $\eta$ is the path-loss exponent,
and b) the linear model
\begin{equation}
P_{rx}\left(z^{\rss}\right)=P_{0}-\rho\, z^{\rss}\label{eq:lineardistance_path_loss}
\end{equation}
In evaluating the regression curves, the values of the parameters
$P_{0}$, $\eta$ and $d_{0}$ ($P_{0}$ and $\rho$) have been optimized
in the first (second) case. The results illustrated in Fig. \ref{fig:rss_fitting_169mhz}
evidence that, as already mentioned in some previous work (e.g., see
\cite{Yang2009} and \cite{Porrat2004}), in indoor LOS conditions
a simple linear law may offer a better match with experimental data
than the standard model (\ref{eq:logdistance_path_loss}); this can
be partly related to wave guiding effects characterizing indoor propagation
along corridors (e.g., see \cite{Bultitude1987,Leung2009,Porrat2004,Soderman2012}).
Actually, the model (\ref{eq:logdistance_path_loss}) might offer
a better match with experimental data if additional RSS measurements
referring to distances greater than $50\meter$ were available; note,
however, that in the considered indoor environment $50\meter$ is
the maximum LOS link distance. For these reasons, in the following,
the model (see (\foreignlanguage{english}{\ref{eq:lineardistance_path_loss}}))
$\psi^{\rss}\left(P_{rx,(m,i)}\right)=\left(P_{0}-P_{rx,(m,i)}\right)/\rho$
is adopted for the acquired RSS measurements; the values selected
for its parameters are listed in Table \ref{tab:statistical_model_summary}.

\subsection{Comparison with Other RSS Experimental Data\label{sub:experimental_comparison}}

In comparing our measurements at $2.4\ghz$ (not shown here for space
limitations) with those referring to $169\mhz$ the following relevant
differences were found: a) a stronger attenuation due to obstructions
(and, in particular, an attenuation increasing faster with $N_{o}$)
is experienced at $2.4\ghz$; b) the log-distance path loss model
(\ref{eq:logdistance_path_loss}) is more accurate than the linear
one (\ref{eq:lineardistance_path_loss}) at $2.4\ghz$; c) a lower
sensitivity to the presence of people moving around the measurement
sites is found at $169\mhz$, since the corresponding wavelength $\lambda_{169\mhz}\simeq1.77$
m is comparable or larger than the typical size of a person; d) spatial
variations in the RSS are slower at $169\mhz$ than those experienced
at $2.4\ghz$ (this results in an increased RSS stability for small
displacements of the antennas of the employed radio devices). These
results suggest that localization systems operating at low frequencies
require a smaller number of anchors (see point a) and exhibit an higher
stability, in a practical scenario (see point d) than the widely used
counterparts exploiting $2.4\ghz$ radios; for this reason, in this
manuscript we focus on measurements acquired at $169\mhz$. Finally,
it is also worth mentioning that very few studies about indoor low-frequency
localization are available in the technical literature \cite{Reynolds2003,Power2009}.

\subsection{TOA Measurements Database\label{sub:toa_measurements}}

\begin{figure}
\centering{}\centering
\def\svgwidth{7cm}
\executeiffilenewer{fig5.svg}{fig5.pdf}%
{inkscape  -z  -D  --file=fig5.svg  %
--export-pdf=fig5.pdf  --export-latex}%
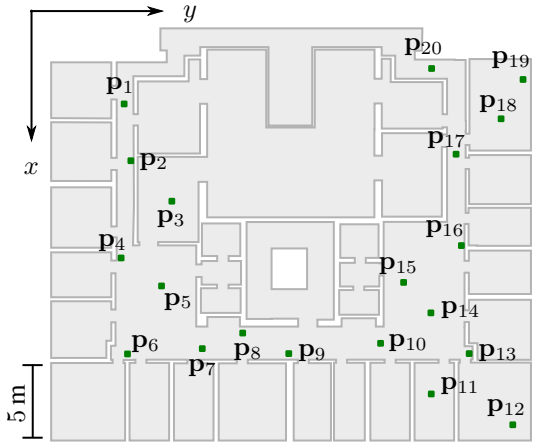%
\caption{Map support $\mathcal{R}$ of the floor where TOA measurements have
been acquired. The $N_{pos}^{\toa}$ measurement sites are indicated
by green squares.\label{fig:map_deis}}
\end{figure}

A campaign for acquiring TOA measurements has not been necessary,
since various experimental data are already publicly available in
the Newcom++ EU database \cite{wprb_database}, whose measurements
refer to $N_{pos}^{\toa}=20$ different sites of the building floor
shown in Fig. \ref{fig:map_deis} and have been acquired using a couple
of Timedomain PulseON 220 \ac{UWB} radios (resulting in 35 NLOS
links and 33 LOS links). The Newcom++ database \cite{wprb_database}
has already been analysed in \cite{Closas2010a,Dardari2010a,Gholami2010};
in particular, in \cite{Closas2010a} it has been shown that its range
measurements cannot be described by a Gaussian model if the bias is
not accounted for. In Appendix \ref{app:mapaware_fitting} it is shown
instead that, if the bias model described in Sec. \ref{sub:observation_model}
is adopted, a good fit can be provided by the Gaussian map-aware model
(\ref{eq:likelihood_with_maps}). 

\subsection{Map-Aware and Map-Unaware Models\label{sub:final_toa_rss_models}}

\begin{figure}
\centering
\def\svgwidth{8cm}
\executeiffilenewer{fig6.svg}{fig6.pdf}%
{inkscape  -z  -D  --file=fig6.svg  %
--export-pdf=fig6.pdf  --export-latex}%
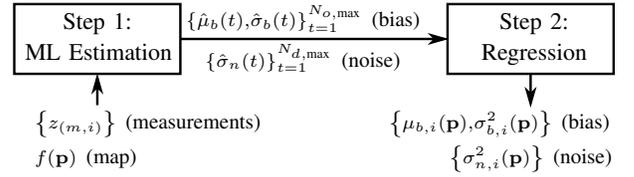%

\caption{Flowchart of the procedure adopted in generating \ac{LS} regression
fits in a map-aware scenario (details in Appendix \ref{app:mapaware_fitting}).\label{fig:fitting}}
\end{figure}

The procedure we adopted in processing the available RSS and TOA measurements
for map-aware fitting consists of the following steps (see Fig. \ref{fig:fitting}):
\begin{enumerate}
\item Randomly partitioning the acquired measurements into a \emph{training}
set, containing \textasciitilde{}25\% of the available experimental
measurements, and a \emph{validation} set, collecting the remaining
part of the measurements; the last set has been processed (see paragraph
\ref{sub:accuracy_test}) to validate the proposed models against
statistically independent data \cite{Bishop2006};
\item Processing map information and the measurements contained in the \emph{training}
set in order to extract a) \ac{ML} estimates of the bias mean and
variance versus the number of obstructions and b) ML estimates of
the noise standard deviation versus the link distance.
\item Developing smooth parametric functions $\ensuremath{\left\{ \mu_{b,i}^{2}(\mathbf{p}),\sigma_{b,i}^{2}(\mathbf{p})\right\} }$
for the bias and a smooth parametric function $\ensuremath{\left\{ \sigma_{n,i}^{2}(\mathbf{p})\right\} }$
for the noise variance by evaluating standard \ac{LS} regression
fits (based on the ML estimates evaluated in the first step). 
\end{enumerate}
In Appendix \ref{app:mapaware_fitting} a detailed description of
these steps is provided; in this paragraph we limit to summarising
the main results. The \emph{map-aware} \emph{likelihood} model we
propose for TOA measurements is\foreignlanguage{english}{
\begin{equation}
\begin{split}f(\mathbf{z}|\tilde{\mathbf{p}})=\prod_{i\in\mathcal{Z}}\mathcal{N} & \left(z_{i};d_{i}(\tilde{\mathbf{p}})+t_{w}N_{o}\left(\tilde{\mathbf{p}},\mathbf{p}_{i}^{a}\right)\sqrt{\epsilon_{r,w}-1},\vphantom{a^{\beta^{\toa}}}\right.\\
 & \:\:\left(\sigma_{b,0}^{\toa}\right)^{2}\left(N_{o}\left(\tilde{\mathbf{p}},\mathbf{p}_{i}^{a}\right)\right)^{2\beta_{b}^{\toa}}+\\
 & \:\:\left.\left(\sigma_{n,0}^{\toa}\right)^{2}\left(d_{i}(\tilde{\mathbf{p}})/d_{0}\right)^{2\beta_{n}^{\toa}}\right)
\end{split}
\label{eq:likelihood_with_maps_toa_final}
\end{equation}
}whereas the \emph{map-aware likelihood} model for RSS measurements
is
\begin{equation}
\begin{split}f(\mathbf{z}|\tilde{\mathbf{p}})=\prod_{i\in\mathcal{Z}}\mathcal{N} & \left(z_{i};d_{i}(\tilde{\mathbf{p}})+\left(\mu_{b,0}^{\rss}+\mu_{b,m}^{\rss}N_{o}\left(\mathbf{p},\mathbf{p}_{i}^{a}\right)\right)\step_{N_{o}},\vphantom{a^{\beta^{\toa}}}\right.\\
 & \:\:\max\left\{ \left(\sigma_{b,0}^{\rss}-\sigma_{b,m}^{\rss}N_{o}\left(\tilde{\mathbf{p}},\mathbf{p}_{i}^{a}\right)\right)^{2},0\right\} +\\
 & \:\:\left.\left(\sigma_{n,0}^{\rss}\right)^{2}\left(d_{i}(\tilde{\mathbf{p}})/d_{0}\right)^{2\beta_{n}^{\rss}}\right)
\end{split}
\label{eq:likelihood_with_maps_rss_final}
\end{equation}
where $\step_{N_{o}}\triangleq\step\left(N_{o}\left(\mathbf{p},\mathbf{p}_{i}^{a}\right)\right)$
. It is important to point out that: a) map-awareness is embedded
in these likelihoods through the $N_{o}\left(\mathbf{p},\mathbf{p}_{i}^{a}\right)$
function; b) the mathematical structure of the bias and noise models
contained in the last two expressions is based on experimental evidence,
as illustrated in Appendix \ref{app:mapaware_fitting}; c) although
simple models can be developed for fitting the mean of the NLOS bias
(see Fig. \ref{fig:rss_fitting_bias_mean}), the same does not hold
for both bias variance and noise variance; d) anomalies in bias variance
and noise variance fitting can be related to various propagation mechanisms
(e.g., reflections and diffractions) not accounted for by the proposed
models. 
\begin{table*}
\begin{centering}

{
\renewcommand{\arraystretch}{1.5}       

{\scriptsize
\begin{tabular}{c|c|ccc|cc}
\toprule
Technology & $\psi(\cdot)$ mapping parameters & \multicolumn{3}{|c|}{Map-aware model parameters} & \multicolumn{2}{c}{Map-unaware model parameters} \\
           &                                  & $\mu_{b,i}(\mathbf{p})$ & $\sigma_{b,i}(\mathbf{p})$ & $\sigma_{n,i}(\mathbf{p})$ & Bias & $\sigma_{n,i}(\mathbf{p})$ \\
\hline
\multirow{2}{*}{TOA UWB}  & \multirow{2}{*}{none ($c_0$ is light speed)} & $\epsilon_{r,w}=5.12$ & $\sigma^{\toa}_{b,0}=0.31\meter$ & $\sigma^{\toa}_{n,0}=0.19\meter$ & \multirow{2}{*}{$\nu^{\toa}_{b}=1.58\meter$} & $\sigma^{\toa}_{n,0}=0.12\meter$ \\
       &      & ($t_{w}=0.35\meter$) & $\beta^{\toa}_b=1.14$ & $\beta^{\toa}_n=0.18$ &       & $\beta^{\toa}_n=0.1$ \\
\hline
\multirow{2}{*}{RSS $169\mhz$} & $\rho=-0.79\dbmm$  & $\mu^{\rss}_{b,0}=12.6\meter$ & $\sigma^{\rss}_{b,0}=7.07\meter$ & $\sigma^{\rss}_{n,0}=2.47\meter$ & $\kappa^{\rss}_{b}=21\meter$ & $\sigma^{\rss}_{n,0}=4.47\meter$ \\
      & $P_0=-35.4\dbm$ & $\mu_{b,m}^{\rss} = 2.53\meter$ & $\sigma^{\rss}_{b,m}=3.0\meter$ & $\beta^{\rss}_n=0.21$ & $\gamma^{\rss}_{b}=2.81\meter$ & $\beta^{\rss}_n= 0.19$ \\
\bottomrule
\end{tabular}
}
}

\par\end{centering}

\centering{}\caption{Numerical values of the parameters appearing in the likelihood functions
(\ref{eq:likelihood_with_maps_toa_final}), (\ref{eq:likelihood_with_maps_rss_final}),
(\ref{eq:likelihood_no_maps_toa_final_exp}), (\ref{eq:likelihood_no_maps_rss_final_gaussian}).
\label{tab:statistical_model_summary}}
\end{table*}

\begin{figure}
\centering{}\includegraphics[width=7cm]{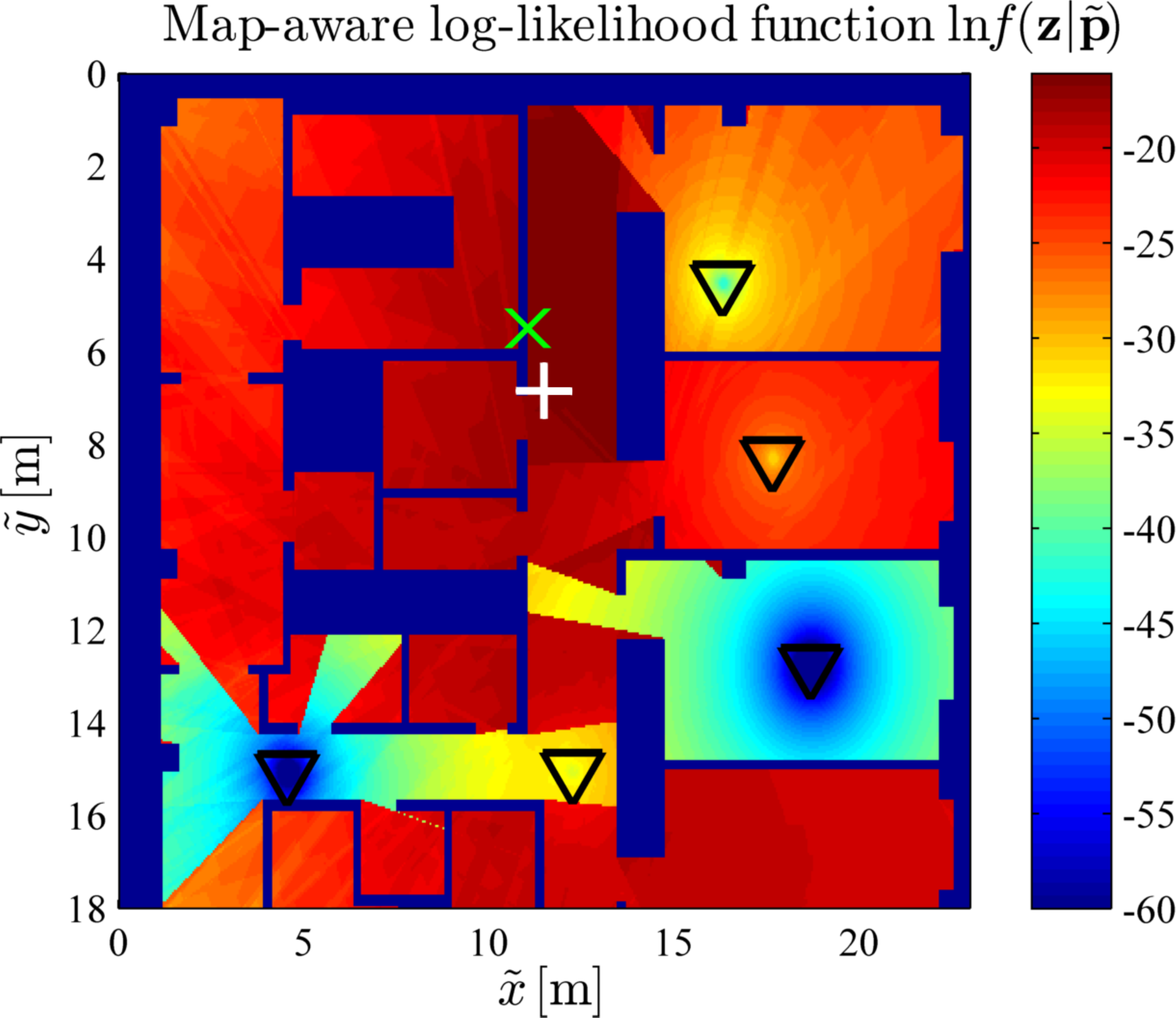}\caption{Representation of the map-aware log-likelihood $\ln f(\mathbf{z}|\tilde{\mathbf{p}}=[\tilde{x};\tilde{y}]^{T})$
(\ref{eq:likelihood_with_maps_rss_final}) evaluated for the specific
map shown in Fig. \ref{fig:map_dii} (the parameter values listed
in Table \ref{tab:statistical_model_summary} for RSS radios have
been employed) and for the specific observation vector $\mathbf{z}$
corresponding to the measurement site \#6 (identified by a white plus
sign), which represents the real agent position. The 5 measurement
sites selected as anchors are identified by black triangles and the
peak of the log-likelihood is evidenced by a green cross (and is $1.4\meter$
away from $\mathbf{p}$).\label{fig:example_log_likelihood_mapaware}}
\end{figure}

\begin{figure}
\centering{}\includegraphics[width=7cm]{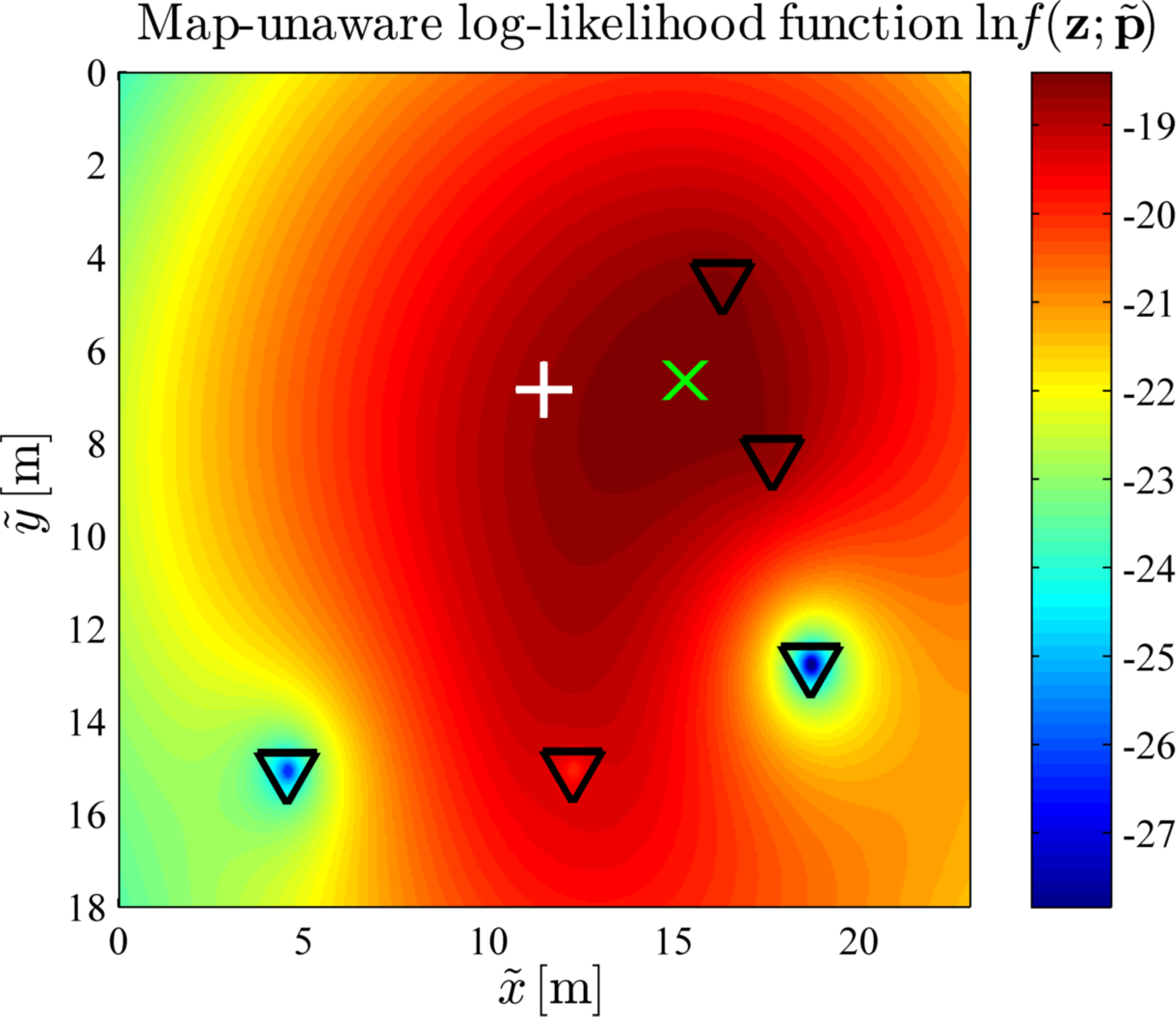}\caption{Representation of the map-unaware log-likelihood $\ln f(\mathbf{z};\tilde{\mathbf{p}}=[\tilde{x};\tilde{y}]^{T})$
(\ref{eq:likelihood_no_maps_rss_final_gaussian}) referring to the
same scenario as Fig. \ref{fig:example_log_likelihood_mapaware},
with $\hat{\mathcal{Z}}^{\nlos}\equiv\mathcal{Z}^{\nlos}$. The 5
measurement sites selected as anchors are identified by black triangles
and the peak of the log-likelihood is evidenced by a green cross (and
is $2.5\meter$ away from $\mathbf{p}$). \label{fig:example_log_likelihood_nomap}}
\end{figure}

As far as map-unaware modelling is concerned, in Appendix \ref{app:nomap_fitting}
a detailed description of the procedure for extracting data fits in
this case is illustrated. Based on the results provided by this procedure,
the \emph{map-unaware likelihood} model for the TOA measurements 
\begin{align}
f(\mathbf{z};\tilde{\mathbf{p}})= & \prod_{i\in\hat{\mathcal{Z}}^{\los}}\mathcal{N}\left(z_{i};d_{i}(\tilde{\mathbf{p}}),\left(\sigma_{n,0}^{\toa}\right)^{2}\left(d_{i}(\tilde{\mathbf{p}})/d_{0}\right)^{2\beta_{n}^{\toa}}\right)\nonumber \\
 & \prod_{i\in\hat{\mathcal{Z}}^{\nlos}}\mathcal{E}\left(z_{i};d_{i}(\tilde{\mathbf{p}}),\nu_{b}^{\toa}\right)\label{eq:likelihood_no_maps_toa_final_exp}
\end{align}
and the \emph{map-unaware likelihood} model for RSS measurements 
\begin{equation}
\begin{split}f(\mathbf{z};\tilde{\mathbf{p}})= & \prod_{i\in\hat{\mathcal{Z}}^{\los}}\mathcal{N}\left(z_{i};d_{i}(\tilde{\mathbf{p}}),\left(\sigma_{n,0}^{\rss}\right)^{2}\left(d_{i}(\tilde{\mathbf{p}})/d_{0}\right)^{2\beta_{n}^{\rss}}\right)\\
 & \prod_{i\in\hat{\mathcal{Z}}^{\nlos}}\mathcal{N}\left(z_{i};d_{i}(\tilde{\mathbf{p}})+\kappa_{b}^{\rss},\vphantom{a^{\beta^{\toa}}}\right.\\
 & \qquad\qquad\left.\left(\sigma_{n,0}^{\rss}\right)^{2}\left(d_{i}(\tilde{\mathbf{p}})/d_{0}\right)^{2\beta_{n}}+\left(\gamma_{b}^{\rss}\right)^{2}\right)
\end{split}
\label{eq:likelihood_no_maps_rss_final_gaussian}
\end{equation}
have been proposed. Estimates of all the parameters appearing in the
likelihood models (\ref{eq:likelihood_with_maps_toa_final})-(\ref{eq:likelihood_no_maps_rss_final_gaussian})
are listed in Table \ref{tab:statistical_model_summary}. Note that:
a) the evaluation of the map-aware likelihoods (\ref{eq:likelihood_with_maps_toa_final})-(\ref{eq:likelihood_with_maps_rss_final})
is more computationally demanding than that of their map-unaware counterparts
(\ref{eq:likelihood_no_maps_toa_final_exp})-(\ref{eq:likelihood_no_maps_rss_final_gaussian}));
b) the significant differences between map-aware and map-unaware models
can be evidenced comparing a plot of (\ref{eq:likelihood_with_maps_rss_final})
with that of (\ref{eq:likelihood_no_maps_rss_final_gaussian}) referring
to the same scenario (see Fig. \ref{fig:example_log_likelihood_mapaware}
and \ref{fig:example_log_likelihood_nomap}, respectively). These
results show that the presence of obstructions (and, in particular,
of walls) introduces jump discontinuities in the map-aware likelihood
function; on the contrary, such discontinuities are not visible in
the representation of its map-unaware counterpart. 

Finally, we would like to point out that, in our opinion, a universal
statistical model for RSS and TOA data is unlikely to exist and that
some adjustments may be required in the models (\ref{eq:likelihood_with_maps_toa_final})-(\ref{eq:likelihood_no_maps_rss_final_gaussian})
if different radio devices are employed. However, we also believe
that our modelling approach (starting from (\ref{eq:likelihood_with_maps})-(\ref{eq:likelihood_no_maps_exp})
and then involving the procedures illustrated in the Appendices) is
quite general and is useful for any reader interested in modelling
experimental data acquired in real world localization systems.

\section{Map-Aware and Map-Unaware Localization: Algorithms and Performance\label{sec:accuracy_results}}

In this Section map-aware and map-unaware estimators are formulated
and compared in terms of accuracy. Moreover, a brief analysis of their
computational complexity is provided.

\subsection{Map-Aware Estimation\label{sub:mapbe}}

The map-aware models (\ref{eq:likelihood_with_maps_toa_final}) and
(\ref{eq:likelihood_with_maps_rss_final}) derived in paragraph \ref{sub:final_toa_rss_models}
can be exploited to derive two types of optimal localization algorithms:
the \ac{MAPBE} and the \ac{MMSEE}. However, our computer simulations
have evidenced that these algorithms perform similarly; for this reason
in the following we focus on the \ac{MAPBE} only, since its mathematical
structure is simpler than that of the \ac{MMSEE}. The \ac{MAPBE}
$\hat{\mathbf{p}}_{\map}(\mathbf{z})$ of the agent position $\mathbf{p}$
can be evaluated as $\hat{\mathbf{p}}_{\map}(\mathbf{z})\triangleq\arg\max_{\tilde{\mathbf{p}}}\ln f(\tilde{\mathbf{p}}|\mathbf{z})$
\cite{kay}. Exploiting (\ref{eq:map_pdf}) and the Bayes' rule, this
estimate can be expressed as 
\begin{equation}
\hat{\mathbf{p}}_{\map}(\mathbf{z})=\arg\max_{\tilde{\mathbf{p}}\in\mathcal{R}^{(\mathcal{Z})}}\ln f(\mathbf{z}|\tilde{\mathbf{p}})\label{eq:MAP}
\end{equation}
where $\tilde{\mathbf{p}}$ is the trial agent position, $\mathcal{R}^{(\mathcal{Z})}$
is the search space (\ref{eq:mapaware_anchor_coverage_region}) and
the pdf $f(\mathbf{z}|\tilde{\mathbf{p}})$ is given by and (\ref{eq:likelihood_with_maps_toa_final})
and (\ref{eq:likelihood_with_maps_rss_final}) for TOA-based and RSS-based
systems, respectively. The evaluation of $\hat{\mathbf{p}}_{\map}$
requires solving a constrained \ac{LS} problem since, thanks to
our unified modelling, the likelihood $f(\mathbf{z}|\tilde{\mathbf{p}})$
appearing in (\ref{eq:MAP}) is always expressed as the product of
Gaussian pdfs. However, finding the global solution to this optimization
problem is not easy since its cost function is non-convex. In principle,
to avoid local minima the \ac{POCS} and the \ac{MDS} techniques
could be exploited; however, refined observation models like (\ref{eq:likelihood_with_maps_toa_final})
and (\ref{eq:likelihood_with_maps_rss_final}) hinder their use. It
is also important to note that the cost function appearing in (\ref{eq:MAP})
is non-differentiable because of the discontinuous behaviour of the
functions (\ref{eq:bias_mean_TOA}), (\ref{eq:bias_mean_RSS}), (\ref{eq:bias_variance_TOA})
and (\ref{eq:bias_variance_RSS}). This prevents the \emph{exact}
use of optimization methods based on the gradient/Hessian matrix of
cost functions (e.g., steepest descent variants). For these reasons,
in our computer simulations the MATLAB routine \texttt{fmincon}, employing
sequential quadratic programming and a numerical \emph{approximation}
of the Hessian matrix, has been employed to solve the problem (\ref{eq:MAP}).

\subsection{Map-Unaware Estimation\label{sub:mle}}

The optimal map-unaware estimator is the \ac{MLE} \cite{kay}.
The \ac{MLE} $\hat{\mathbf{p}}_{\ml}(\mathbf{z})$ of $\mathbf{p}$
can be expressed as \cite{kay} 
\begin{equation}
\hat{\mathbf{p}}_{\ml}(\mathbf{z})=\arg\max_{\tilde{\mathbf{p}}\in\mathcal{P}^{(\mathcal{Z})}}\ln f(\mathbf{z};\tilde{\mathbf{p}})\label{eq:ML}
\end{equation}
where $\tilde{\mathbf{p}}$ is the trial agent position, $\mathcal{P}^{(\mathcal{Z})}$
is the search space (\ref{eq:nomap_anchor_coverage_region}) and the
pdf $f(\mathbf{z};\tilde{\mathbf{p}})$ is given by (\ref{eq:likelihood_no_maps_toa_final_exp})
and (\ref{eq:likelihood_no_maps_rss_final_gaussian}) for TOA-based
and RSS-based systems, respectively. Note that the cost function characterising
the \ac{MLE} (\ref{eq:ML}) is discontinuous and non-linear because
of the assumption of a LOS/NLOS detection in identifying $\hat{\mathcal{Z}}^{\los}$
and $\hat{\mathcal{Z}}^{\nlos}$. In addition, in the TOA case, the
use of exponential pdfs for modelling NLOS links (see (\ref{eq:likelihood_with_maps_toa_final}))
entails that $f(\mathbf{z};\tilde{\mathbf{p}})=0$ for a large portion
of the domain $\mathcal{P}^{(\mathcal{Z})}$ and this complicates
the search for global minima. A simple approximation that may be employed
to mitigate this problem is to replace the function $\mathcal{E}\left(r;t,\mu\right)$
appearing in the right-hand side of (\ref{eq:likelihood_no_maps_toa_final_exp})
with the pdf mixture $\mathcal{M}(r;t,\mu)\triangleq0.8\cdot\mathcal{E}\left(r;t,\mu\right)+0.2\cdot\mathcal{N}\left(r;t,\frac{\mu^{2}}{100}\right)\step(t-r)$,
which exhibits a decreasing exponential behaviour for $r\geq t$ and
a small Gaussian tail for $r<t$. Moreover, in order to further mitigate
the problem of local minima, in our work the search domain $\mathcal{P}^{(\mathcal{Z})}$
has been partitioned into $N_{sub}$ sub-rectangles (whose sides have
a length not exceeding $10\meter$); then, the MATLAB \texttt{fmincon}
routine has been run $N_{sub}$ times over each of the sub-rectangles
and the final \ac{MLE} has been selected on the basis of a minimum-cost
criterion.

\subsection{Accuracy of Map-Aware and Map-Unaware Localization\label{sub:accuracy_test}}

\begin{figure}
\centering{}\centering
\setlength\figureheight{5cm}
\setlength\figurewidth{7cm}
\input{fig9.tikz}\caption{RMSE performance of the map-aware estimator (\ref{eq:MAP}) and the
map-unaware estimator (\ref{eq:ML}), for two different values of
$P_{e}^{\nlos}$. RSS-based localization is considered.\label{fig:model_comparison_rss}}
\end{figure}
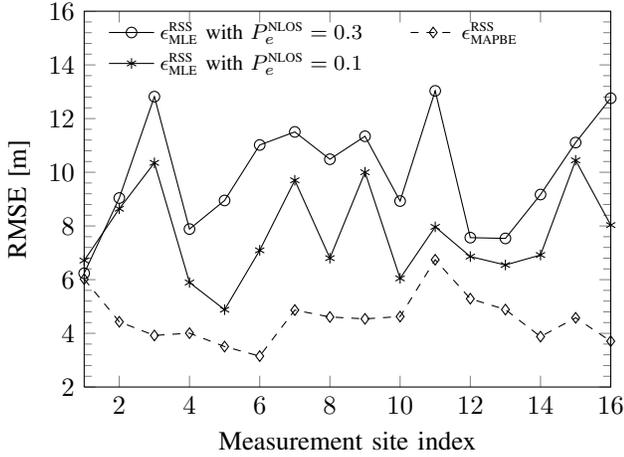
\begin{figure}
\begin{centering}
\centering
\setlength\figureheight{5cm}
\setlength\figurewidth{7cm}
\input{fig10.tikz}
\par\end{centering}

\centering{}\caption{RMSE performance of the map-aware estimator (\ref{eq:MAP}) and the
map-unaware estimator (\ref{eq:ML}), for $P_{e}^{\nlos}=0.1$. TOA-based
localization is considered.\label{fig:model_comparison_toa}}
\end{figure}
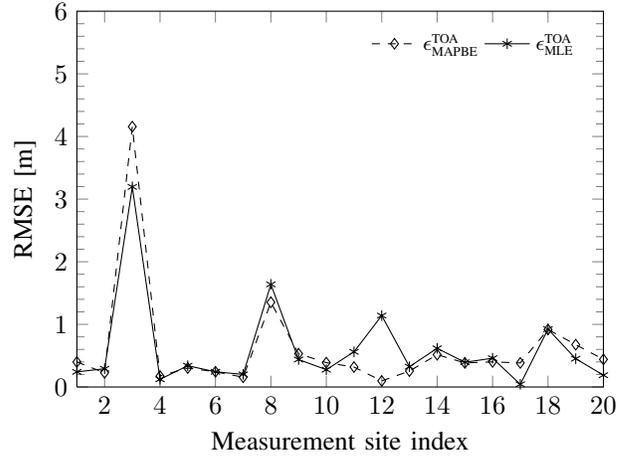

The accuracy provided by the estimators (\ref{eq:MAP}) and (\ref{eq:ML})
has been assessed by means of a simulation tool implementing the following
procedure: 
\begin{enumerate}
\item A measurement site, identified by the index $m_{a}$, is randomly
selected in the set $\left\{ 1,2,...,N_{pos}\right\} $ referring
to the measurement sites shown in Fig. \ref{fig:map_dii} (RSS case)
or in Fig. \ref{fig:map_deis} (TOA case). The position of this site
represents the true position $\mathbf{p}$ of the agent, so that $\mathbf{p}=\mathbf{p}_{m_{a}}$.
\item For the selected agent position, in the validation set introduced
in Sec. \ref{sub:final_toa_rss_models}, $N_{obs}^{V}\left(m_{a}\right)$
sets of measurements associated with $N_{obs}^{V}\left(m_{a}\right)$
distinct sites are available. Each of these sites can be employed
as a ``virtual'' anchor, and in each selected set (containing $N_{z}^{\rss}$
or $N_{z}^{\toa}$ measurements) one measurement is randomly selected
and placed in the observation vector $\mathbf{z}$. To make our simulation
results realistic, if $N_{obs}^{V}\left(m_{a}\right)>5$, only the
observations associated with 5 random sites are chosen.
\item The estimates $\hat{\mathbf{p}}_{\map}(\mathbf{z})$ (\ref{eq:MAP})
and $\hat{\mathbf{p}}_{\ml}(\mathbf{z})$ (\ref{eq:ML}) and the associated
\emph{localization errors} $\left\Vert \mathbf{p}-\hat{\mathbf{p}}_{\map}(\mathbf{z})\right\Vert $
and $\left\Vert \mathbf{p}-\hat{\mathbf{p}}_{\ml}(\mathbf{z})\right\Vert $
are evaluated. At the end of the simulation such errors are processed
to assess the \ac{RMSE} performance.
\item Steps 1-3 are repeated until at least $N_{runs}=2\cdot10^{4}$ iterations
have been carried out.
\end{enumerate}
Some RMSE performance results $\epsilon_{\map}$ and $\epsilon_{\ml}$
(referring to the MAPBE (\ref{eq:MAP}) and the MLE (\ref{eq:ML}),
respectively) are shown in Fig. \ref{fig:model_comparison_rss} and
\ref{fig:model_comparison_toa} for RSS and TOA localization, respectively.
These results evidence that:
\begin{enumerate}
\item In the RSS case the \ac{MAPBE} always outperforms its map-unaware
\ac{MLE} counterpart (on the average $\epsilon_{\ml}^{\rss}=1.7\epsilon_{\map}^{\rss}$
for $P_{e}^{\nlos}=0.1$ and $\epsilon_{\ml}^{\rss}=2.1\epsilon_{\map}^{\rss}$
for $P_{e}^{\nlos}=0.3$; in other words, map-aware modelling enhance
localization accuracy by 70\% for $P_{e}^{\nlos}=0.1$ and by 110\%
for $P_{e}^{\nlos}=0.3$). This is due to the fact that the former
technique mitigates NLOS bias (which represents a major source of
error) more accurately than the latter one. In fact, maps provide
a significant help whenever the bias mean predicted by (\ref{eq:bias_mean_RSS})
is significantly different from $\kappa_{b}^{\rss}$ (which accounts
for NLOS propagation in the \ac{MLE}), i.e. whenever there are
several obstructions between the agent and the anchors (see Fig. \ref{fig:rss_fitting_bias_mean}).
\item In the TOA case map-aware and map-unaware estimators perform similarly
and offer good accuracy; this is due to the fact that the TOA measurements
stored in the database \cite{wprb_database} are not affected by large
NLOS biases (since all such measurements refer to close sites, often
experiencing LOS conditions), so that map awareness does not play
a significant role in this case. Note also that localization errors
for both \ac{MAPBE} and \ac{MLE} are smaller than $1\meter$
for all measurement sites except \#3, \#8 and \#12. This result can
be explained referring to the specific case of site \#3, for which
it has been found that the large error is due to a combination of
a) an anchor placement characterized by an high \emph{geometric dilution
of precision} \cite{Sharp2009} and b) undetected direct paths in
the ranging phase (see \cite{Alavi2006}) resulting in highly biased
measurements. It is also interesting to point out that the \ac{MLE}
is slightly more accurate than the \ac{MAPBE} in the measurement
sites \#3, \#17, \#19 and \#20; such results can be related to propagation
effects not accounted for by our models (e.g., reflections and corridor
waveguiding); however, in these cases, the difference between \ac{MLE}
and \ac{MAPBE} RMSE is very small (few centimetres). 
\item The performance gap between the RSS-based \ac{MAPBE} and the \ac{MLE}
is large ($\epsilon_{\ml}^{\rss}-\epsilon_{\map}^{\rss}=4.9\meter$
on the average), while TOA-based algorithms perform similarly ($\epsilon_{\ml}^{\toa}\simeq\epsilon_{\map}^{\toa}$
on the average). This shows that maps play a significant role in improving
localization accuracy when the quality of available measurements is
poor, i.e., when measurements are affected by significant NLOS bias
and noise (this occurs in RSS-based systems); on the contrary, if
the quality is good (like in TOA UWB systems), accuracy is not significantly
enhanced by map awareness (some theoretical results about this can
be found in \cite{Montorsi_APRIORI_BOUNDS_TIT,Montorsi_APRIORI_IMPACT_ICC_2013}).
\item Map-aware fitting provides good results even if a small number of
links are used in the training set (i.e., in the fitting phase); in
fact, in the RSS case, only 32 links (\textasciitilde{}25\% of the
116 available links) composed the training set and the performance
shown in Fig. \ref{fig:model_comparison_rss} have been obtained using
the validation set.
\end{enumerate}
Our numerical results also confirm some well-known results and, in
particular, that:
\begin{enumerate}
\item RSS-based localization algorithms are substantially outperformed by
their UWB TOA-based counterparts (on the average, $\epsilon_{\map}^{\rss}=4.4\meter$
against $\epsilon_{\map}^{\toa}=0.4\meter$), thanks to the superior
ranging capabilities provided by UWB waveforms.
\item Significant variations in the RMSE of both TOA-based and RSS-based
estimators are found when different sites are considered (similar
results can be found in \cite[Fig. 2]{Gholami2010}, although in the
scenario we consider both \ac{MAPBE} and \ac{MLE} perform better
than the \ac{LS} and \ac{MDS} algorithms analysed in that reference).
This is due, in the TOA case, to errors in TOA estimation originating
from significant multipath \cite{Alavi2006}; in the RSS case, instead,
variations originate from the spatial selectivity experienced in indoor
environments.
\end{enumerate}
Finally, it is worth pointing out that some additional numerical results
(e.g., referring to the robustness of the proposed statistical models
to parametric inaccuracies) have not been included here for space
limitations, but are publicly available online (see \cite{Montorsi_MAPAWARE_MODELS_ICC_2013}).

\subsection{Complexity of Map-Aware and Map-Unaware Localization}

In this Paragraph a brief analysis of the computational complexity
of the \ac{MAPBE} and \ac{MLE} is provided. To begin, we note
that the complexity associated with the evaluation of $N_{o}(\cdot,\cdot)$
is $\mathcal{O}(N_{s})$ if the shape of map obstructions can be approximated
by $N_{s}$ segments, since $N_{s}$ line intersection tests are involved
in such a case. Therefore, the complexity associated with the evaluation
of the map-aware likelihood $f(\mathbf{z}|\tilde{\mathbf{p}})$ (see
(\ref{eq:likelihood_with_maps_toa_final})-(\ref{eq:likelihood_with_maps_rss_final}))
is approximately $\mathcal{O}\left(\bar{N}_{s}\cdot N_{obs}\right)$,
where $\bar{N}_{s}$ is the average number of segment intersection
tests accomplished in processing each of the $N_{obs}$ observations%
\footnote{Note that $\bar{N}_{s}$ is non-linearly related to the shape of obstructions,
to the trial agent position and to the anchor positions.%
}. Then, it can be inferred that the overall computational complexity
of the \ac{MAPBE} (\ref{eq:MAP}) is $\mathcal{O}\left(\bar{N}_{s}\cdot N_{obs}\cdot N_{eval}^{\map}\right)$,
where $N_{eval}^{\map}$ denotes the overall number of times the likelihood
function (\ref{eq:likelihood_with_maps_toa_final}) or (\ref{eq:likelihood_with_maps_rss_final})
is computed. Unluckily, the parameter $N_{eval}^{\map}$ cannot be
easily related to the other parameters of the proposed \ac{MAPBE}
strategy (\ref{eq:MAP}), since it is non-linearly related to the
shape of $f(\mathbf{z}|\tilde{\mathbf{p}})$ (which is non-convex
and non-differentiable) and depends strongly on the exact type of
optimization method employed (see Paragraph \ref{sub:mapbe}).

As far as the map-unaware \ac{MLE} (\ref{eq:ML}) is concerned,
its overall complexity can be expressed as $\mathcal{O}\left(N_{obs}\cdot N_{eval}^{\ml}\right)$
since the complexity associated with the evaluation of the likelihood
$f(\mathbf{z};\tilde{\mathbf{p}})$ (\ref{eq:likelihood_no_maps_toa_final_exp})-(\ref{eq:likelihood_no_maps_rss_final_gaussian})
is $\mathcal{O}\left(N_{obs}\right)$; note that $N_{eval}^{\ml}$
denotes the overall number of times the likelihood function (\ref{eq:likelihood_no_maps_toa_final_exp})
or (\ref{eq:likelihood_no_maps_rss_final_gaussian}) is evaluated
and, just like $N_{eval}^{\map}$, it cannot be easily related to
the other parameters of the \ac{MLE}. The constraint $\tilde{\mathbf{p}}\in\mathcal{P}^{(\mathcal{Z})}$
appearing in (\ref{eq:ML}) makes even more difficult obtaining a
closed expression for $N_{eval}^{\ml}$; however, it cannot be neglected,
specially on large maps, because: a) the map-unaware likelihood functions
(\ref{eq:likelihood_no_maps_toa_final_exp})-(\ref{eq:likelihood_no_maps_rss_final_gaussian})
are not defined for $\tilde{\mathbf{p}}\notin\mathcal{P}^{(\mathcal{Z})}$;
b) even if the likelihood domains are extended to $\mathbb{R}^{2}$,
minima different from the global one are substantially less likely
to be contained in the domain $\mathcal{P}^{(\mathcal{Z})}$ than
in the whole $\mathbb{R}^{2}$ space. For these reasons, constrained
optimization methods capable of handling the constraint $\tilde{\mathbf{p}}\in\mathcal{P}^{(\mathcal{Z})}$
must be selected (eventually employing heuristic approximations; see
Paragraph \ref{sub:mle}). 

The considerations illustrated above led us to the conclusion that
an accurate estimation of the computational complexity of both (\ref{eq:MAP})
and (\ref{eq:ML}) requires a mixed analytical/simulative approach;
for more details, the reader is referred to \cite{Montorsi_REDUCED_COMPLEXITY},
which provides an in-depth computational complexity analysis together
with reduced-complexity algorithms.

\section{Conclusions\label{sec:conclusions}}

In this manuscript novel statistical models for map-aware indoor localization
have been developed. Such models are based on experimental evidence
and, in particular, rely on TOA UWB and low-frequency narrowband RSS
measurements acquired in experimental campaigns. These measurements
have been exploited to a) develop suitable parametric models for NLOS
bias and noise, b) estimate the values of the parameters appearing
in such models and c) assess the accuracy of map-aware localization
techniques based on the proposed models. Our results have evidenced
that map-aware localization techniques may significantly outperform
their optimal map-unaware counterparts, even when state-of-the-art
map-unaware algorithms based on LOS/NLOS detection and mitigation
are considered. Finally, we would like to point out that:
\begin{enumerate}
\item The experimental campaign needed to estimate the parameters of the
proposed model can be accomplished much more easily than those commonly
required by fingerprinting methods.
\item The likelihoods (\ref{eq:likelihood_with_maps_toa_final}) and (\ref{eq:likelihood_with_maps_rss_final})
and the parameters listed in Table \ref{tab:statistical_model_summary}
can be exploited by localization algorithms operating in environments
different from the ones considered here, provided that similar conditions
(in terms of bandwidth of localization signals, wall thickness and
composition, etc.) are experienced. 
\item Even if the case of a \emph{static} agent has been taken into consideration
in this manuscript, the application of the developed models to navigation
systems (where a \emph{mobile }agent is tracked) is straightforward;
for instance, (\ref{eq:likelihood_with_maps}) can be adopted as a
\emph{measurement model} for tracking filters (e.g., Kalman or particle
filters). 
\end{enumerate}

\appendix

\section{Procedures for Fitting Measurements}

\subsection{Fitting Measurements in the Map-Aware Case\label{app:mapaware_fitting}}

In this paragraph the two steps of the fitting procedure (see Fig.
\ref{fig:fitting} and paragraph \ref{sub:final_toa_rss_models})
leading to the models (\ref{eq:likelihood_with_maps_toa_final}) and
(\ref{eq:likelihood_with_maps_rss_final}) and to the parameter values
listed in Table \ref{tab:statistical_model_summary} are described.

\subsubsection{ML Estimation}

The first step of the above mentioned procedure includes three sub-procedures:
data modelling, bias estimation and noise estimation.

\emph{Data modelling} - Let $(m,i)\in\mathcal{I}$ denote the link
between the $m$-th and the $i$-th measurement sites, where $\mathcal{I}$
denotes the the set of the couples of indices identifying each available
link. We assume that the residual rvs $\{r_{(m,i)}\triangleq z_{(m,i)}-d_{(m,i)}\}$
follow a Gaussian distribution (see the proposed data model (\ref{eq:likelihood_with_maps}))
\begin{equation}
f\left(\left\{ r_{(m,i)}\right\} \right)=\mathcal{N}\left(r_{(m,i)};\mu_{b,i}(\mathbf{p}_{m}),\sigma_{b,i}^{2}(\mathbf{p}_{m})+\sigma_{n,i}^{2}(\mathbf{p}_{m})\right)\label{eq:pdf_single_measurement}
\end{equation}
The validity of the Gaussian assumption (\ref{eq:pdf_single_measurement})
has been analysed resorting to the Anderson-Darling normality test
\cite{Closas2010a,Anderson1952}, which relies on the computation
of a statistic, commonly denoted $A^{2*}$, and on its comparison
with a fixed threshold $\gamma_{A}=1.159$ (if a significance level
equal to $0.05$ is selected%
\footnote{The lower the selected significance level, the stronger should be
the evidence required to pass the test. The choice of the significance
level is somewhat arbitrary; however, a level equal to 5\% is usually
selected for various applications.%
}). We computed the $A^{2*}$ statistic for the sets of measurements
$\left\{ r_{(m,i)}\right\} $ referring to the RSS radios and associated
with the links crossing $1$, $2$, $3$, $4$, 5, 6 walls%
\footnote{There were not enough data to run the normality test in the other
cases.%
} and for these sets it was found that $A^{2*}$ is equal to $0.48$,
$0.24$, $0.26$, $0.25$, $0.20$ and $0.42$, respectively; therefore,
the Gaussian assumption (\ref{eq:pdf_single_measurement}) can be
deemed valid in all the considered cases.

Then, the joint pdf of the experimental measurements $\ensuremath{\left\{ z_{(m,i)}\right\} }$
can be put in a useful form if:
\begin{enumerate}
\item Statistical independence among different links is assumed.
\item The RSS bias mean $\mu_{b,i}^{\rss}(\mathbf{p}_{m})$ and RSS/TOA/TDOA
bias variance $\sigma_{b,i}^{2}(\mathbf{p}_{m})$ are assumed to depend
on the number of obstructions $t_{o,(m,i)}\triangleq N_{o}\left(\mathbf{p}_{m},\mathbf{p}_{i}\right)$
associated with the link $(m,i)$, so that $\mu_{b,i}(\mathbf{p}_{m})$
and $\sigma_{b,i}(\mathbf{p}_{m})$ can be denoted $\mu_{b}\left(t_{o,(m,i)}\right)$
and $\sigma_{b}\left(t_{o,(m,i)}\right)$, respectively. Note that
$t_{o,(m,i)}$ is known at this stage.
\item The noise term $\sigma_{n,i}^{2}(\mathbf{p}_{m})$ is assumed to depend
on the distance $d_{(m,i)}\triangleq d_{i}(\mathbf{p}_{m})$; such
a dependence can be assessed estimating the function values $\left\{ \sigma_{n,i}\right\} $
associated with $N_{d,\mymax}\triangleq\max_{i}\left\lfloor d_{\mymax,i}/\triangle\right\rfloor $
``distance bins''; here $\triangle$ represents the distance bin
size (in meters) and $d_{\mymax,i}$ is the maximum connectivity distance
for the $i$-th anchor (see Sec. \ref{sub:connectivity_model}). Then,
a staircase approximation for each of the continuous functions $\left\{ \sigma_{n,i}(\cdot)\right\} $
can be adopted; the values associated with the steps are represented
by the set $\left\{ \sigma_{n}\left(t\right)\right\} _{t=1}^{N_{d,\mymax}}$,
where $\sigma_{n}\left(t\right)$ is the noise standard deviation
referring to the distance values that belong to the distance bin $[t\triangle;(t+1)\triangle]$.
For this reason, $\sigma_{n,i}(\mathbf{p}_{m})$ can be denoted $\sigma_{n}\left(t_{d,(m,i)}\right)$,
where $t_{d,(m,i)}$ is the index of the distance bin associated with
the distance $d_{(m,i)}$ of the link $(m,i)$ and is known at this
stage.
\end{enumerate}
In fact, if the assumptions listed above hold, the joint pdf of $\ensuremath{\left\{ r_{(m,i)}\right\} }$
(\ref{eq:pdf_single_measurement}) can be expressed as
\begin{equation}
\begin{split}f\left(\left\{ r_{(m,i)}\right\} \right)=\prod_{(m,i)\in\mathcal{I}}\mathcal{N} & \left(r_{(m,i)};\mu_{b}\left(t_{o,(m,i)}\right),\right.\\
 & \:\left.\sigma_{b}^{2}\left(t_{o,(m,i)}\right)+\sigma_{n}^{2}\left(t_{d,(m,i)}\right)\right)
\end{split}
\label{eq:pdf_all_measurements}
\end{equation}
\emph{Bias estimation} - Estimates $\ensuremath{\left\{ \hat{\mu}_{b}(t),\hat{\sigma}_{b}(t)\right\} }$
of the parameters $\ensuremath{\left\{ \mu_{b}\left(t_{o,(m,i)}\right),\sigma_{b}\left(t_{o,(m,i)}\right)\right\} }$
of (\ref{eq:pdf_all_measurements}) can now be evaluated. This requires
processing all the measurements $\ensuremath{\{z_{(m,i)}\}}$ such
that $t_{o,(m,i)}=t$, i.e., the set $\left\{ r_{(m,i)}|(m,i)\in\mathcal{I}_{b}(t)\right\} $,
with $\mathcal{I}_{b}(t)\triangleq\left\{ (m,i)|t_{o,(m,i)}=t\right\} $,
since this set represents a sufficient statistic%
\footnote{It is easy to proof the sufficiency of $\left\{ z_{(m,i)}|(m,i)\in\mathcal{I}_{b}(n)\right\} $
for the estimation of $\mu_{b}\left(n\right)$ using the Neyman-Fisher
theorem \cite[Sec. 5.4]{kay} and assuming that $\mu_{b}\left(n\right)$
is a parameter independent from $\left\{ \mu_{b}\left(m\right),\forall m\neq n\right\} $.
This assumption is indeed valid at this stage since only after evaluating
regression fits the random quantities $\left\{ \mu_{b}\left(n\right)\right\} $
are ``linked'' together.%
}. The \ac{MLE} for \foreignlanguage{english}{$\mu_{b}\left(t_{o,(m,i)}\right)$},
given the above mentioned set, is (see (\ref{eq:pdf_all_measurements}))
\begin{align}
\hat{\mu}_{b}(t) & \triangleq\arg\max_{\tilde{\mu}}\ln f\left(\left\{ r_{(m,i)},(m,i)\in\mathcal{I}_{b}(t)\right\} ;\mu_{b}\left(t\right)=\tilde{\mu}\right)\nonumber \\
 & =\frac{1}{\left|\mathcal{I}_{b}(t)\right|}\sum_{(m,i)\in\mathcal{I}_{b}(t)}r_{(m,i)}\label{eq:bias_mean_estimator}
\end{align}
for $t=1,...,N_{o,\mymax}$, where $N_{o,\mymax}$ is the maximum
number of obstructions affecting the links considered in our measurement
campaign; note that the right-hand side of (\ref{eq:bias_mean_estimator})
represents the sample average of the residuals $\left\{ r_{(m,i)}\right\} $.
Similarly, the \ac{MLE} for \foreignlanguage{english}{$\sigma_{b}\left(t_{o,(m,i)}\right)$}
is given by (see (\ref{eq:pdf_all_measurements})) 
\begin{multline}
\hat{\sigma}_{b}(t)\triangleq\arg\max_{\tilde{\sigma}}\ln f\left(\left\{ r_{(m,i)},(m,i)\in\mathcal{I}_{b}(t)\right\} ;\sigma_{b}\left(t\right)=\tilde{\sigma}\right)\\
=\arg\min_{\tilde{\sigma}}\sum_{(m,i)\in\mathcal{I}_{b}(t)}\left\{ \frac{\left(r_{(m,i)}-\mu_{b}\left(t\right)\right)^{2}}{\tilde{\sigma}^{2}+\sigma_{n}^{2}\left(t_{d,(m,i)}\right)}\right.\\
\left.+2\ln\sqrt{\tilde{\sigma}^{2}+\sigma_{n}^{2}\left(t_{d,(m,i)}\right)}\vphantom{\frac{\left(r_{(m,i)}-\mu_{b}\left(t\right)\right)^{2}}{\tilde{\sigma}^{2}+\sigma_{n}^{2}\left(t_{d,(m,i)}\right)}}\right\} \label{eq:bias_var_estimator1}
\end{multline}
for $t=1,...,N_{o,\mymax}$. Unluckily, solving the last problem requires
the knowledge of models for $\mu_{b}\left(t\right)$ and for $\sigma_{n}^{2}\left(t_{d,(m,i)}\right)$.
To circumvent this, we assume that the bias mean $\mu_{b}\left(t\right)$
is well approximated by its estimate $\hat{\mu}_{b}(t)$ and that
$\sigma_{n}\left(t_{d,(m,i)}\right)\ll\sigma_{b}\left(t\right)$,
so that (\ref{eq:bias_var_estimator1}) can be simplified as 
\begin{align}
\hat{\sigma}_{b}(t) & =\arg\min_{\tilde{\sigma}}\sum_{(m,i)\in\mathcal{I}_{b}(t)}\left\{ \left(\frac{r_{(m,i)}-\hat{\mu}_{b}(t)}{\tilde{\sigma}}\right)^{2}+2\ln\tilde{\sigma}\right\} \nonumber \\
 & =\frac{1}{\left|\mathcal{I}_{b}(t)\right|}\sum_{(m,i)\in\mathcal{I}_{b}(t)}\left(r_{(m,i)}-\hat{\mu}_{b}(t)\right)^{2}\label{eq:bias_var_final}
\end{align}
The last result shows that \foreignlanguage{english}{$\hat{\sigma}_{b}(t)$}
is given by a (biased) variance of the residuals $\left\{ r_{(m,i)}\right\} $. 

\emph{Noise estimation} - The estimation of the noise standard deviation
$\sigma_{n}\left(t_{d,(m,i)}\right)$ is based on the set $\left\{ r_{(m,i)}|(m,i)\in\mathcal{I}_{n}(t)\right\} $
with $\mathcal{I}_{n}(t)\triangleq\left\{ (m,i)|t_{d,(m,i)}=t\right\} $,
since this set represents a sufficient statistic. The \ac{MLE}
is (see (\ref{eq:pdf_all_measurements})) 
\begin{align}
\hat{\sigma}_{n}(t) & \triangleq\arg\max_{\tilde{\sigma}}\ln f\left(\left\{ r_{(m,i)},(m,i)\in\mathcal{I}_{n}(t)\right\} ;\sigma_{n}\left(t\right)=\tilde{\sigma}\right)\nonumber \\
 & =\arg\min_{\tilde{\sigma}}\sum_{(m,i)\in\mathcal{I}_{n}(t)}\left\{ \frac{\left(r_{(m,i)}-\mu_{b}\left(t_{o,(m,i)}\right)\right)^{2}}{\sigma_{b}^{2}\left(t_{o,(m,i)}\right)+\tilde{\sigma}^{2}}\right.\nonumber \\
 & \qquad\qquad\qquad\qquad\left.+2\ln\sqrt{\sigma_{b}^{2}\left(t_{o,(m,i)}\right)+\tilde{\sigma}^{2}}\vphantom{\frac{\left(r_{(m,i)}-\mu_{b}\left(t_{o,(m,i)}\right)\right)^{2}}{\sigma_{b}^{2}\left(t_{o,(m,i)}\right)+\tilde{\sigma}^{2}}}\right\} \label{eq:noise_var_estimator1}
\end{align}
for $t=1,...,N_{d,\mymax}$. Like in the previous case, we assume
that the bias mean $\mu_{b}\left(t_{o,(m,i)}\right)$ and variance
$\sigma_{b}^{2}\left(t_{o,(m,i)}\right)$ are well approximated by
their estimates $\hat{\mu}_{b}\left(t_{o,(m,i)}\right)$ and $\hat{\sigma}_{b}\left(t_{o,(m,i)}\right)$,
so that (\ref{eq:noise_var_estimator1}) turns into
\begin{multline}
\hat{\sigma}_{n}(t)=\arg\min_{\tilde{\sigma}}\sum_{(m,i)\in\mathcal{I}_{n}(t)}\left\{ \frac{\left(r_{(m,i)}-\hat{\mu}_{b}\left(t_{o,(m,i)}\right)\right)^{2}}{\hat{\sigma}_{b}^{2}\left(t_{o,(m,i)}\right)+\tilde{\sigma}^{2}}\right.\\
\left.+2\ln\sqrt{\hat{\sigma}_{b}^{2}\left(t_{o,(m,i)}\right)+\tilde{\sigma}^{2}}\vphantom{\frac{\left(r_{(m,i)}-\hat{\mu}_{b}\left(t_{o,(m,i)}\right)\right)^{2}}{\hat{\sigma}_{b}^{2}\left(t_{o,(m,i)}\right)+\tilde{\sigma}^{2}}}\right\} \label{eq:noise_var_estimator_final}
\end{multline}
which, unluckily, cannot be put in a closed form. For this reason,
the last problem has been solved numerically resorting to the MATLAB
\texttt{fmincon} routine.

\subsubsection{Evaluation of Regression Fits\label{app:regression_mapaware}}

In the second step of the procedure illustrated in Fig. \ref{fig:fitting},
the ML estimates acquired in the first step are processed to extract
two \ac{LS} regression fits, one for the bias, the other one for
the noise model.

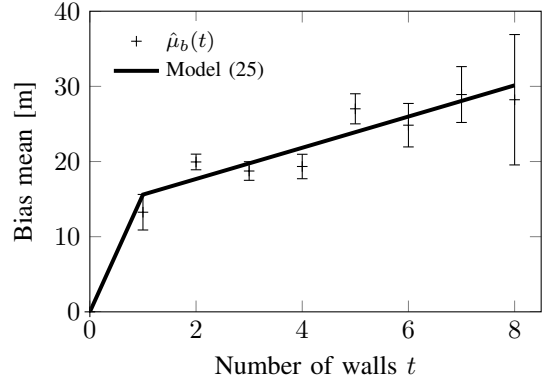
\begin{figure}
\centering
\setlength\figureheight{4cm}
\setlength\figurewidth{6cm}
\input{fig11.tikz}

\caption{Bias mean versus the number of obstructions for RSS measurements.
Note that, for each ML estimate $\hat{\mu}_{b}(t)$ (\ref{eq:bias_mean_estimator}),
the associated error bar is inversely proportional to the number of
links $\left|\mathcal{I}_{b}(t)\right|$ available for its evaluation.\label{fig:rss_fitting_bias_mean}}
\end{figure}

\emph{Bias model fit} - As already discussed in paragraph \ref{sub:observation_model},
in the TOA/TDOA case a simple \emph{linear} regression procedure can
be employed for extracting the average parameters $t_{w}$ and $\epsilon_{r,w}$
appearing in $\mu_{b,i}(\cdot)$ (\ref{eq:bias_mean_TOA}) from the
ML estimates $\ensuremath{\left\{ \hat{\mu}_{b}(t)\right\} _{t=1}^{N_{o,\mymax}}}$.
In the RSS case, instead, the simple linear regression function, based
on experimental evidence (see Fig. \ref{fig:rss_fitting_bias_mean}),
\begin{equation}
\mu_{b,i}^{\rss}(\mathbf{p})=\left(\mu_{b,0}^{\rss}+\mu_{b,m}^{\rss}N_{o}\left(\mathbf{p},\mathbf{p}_{i}^{a}\right)\right)\step_{N_{o}}\label{eq:bias_mean_RSS}
\end{equation}
is proposed; here $\mu_{b,0}^{\rss}$ and $\mu_{b,m}^{\rss}$ are
expressed in meters and the factor $\step_{N_{o}}\triangleq\step\left(N_{o}\left(\mathbf{p},\mathbf{p}_{i}^{a}\right)\right)$
forces $\mu_{b,i}^{\rss}(\mathbf{p})$ (\ref{eq:bias_mean_RSS}) to
zero when $N_{o}\left(\mathbf{p},\mathbf{p}_{i}^{a}\right)=0$. Note
that the selected model differs from the AF bias model, characterized
by a logarithmic dependence on $N_{o}\left(\mathbf{p},\mathbf{p}_{i}^{a}\right)$.
As far as the\emph{ }standard deviation $\sigma_{b,i}(\mathbf{p})$
of the TOA bias is concerned, the power-like model 
\begin{equation}
\sigma_{b,i}^{\toa}(\mathbf{p})=\sigma_{b,0}^{\toa}\cdot\left(N_{o}\left(\mathbf{p},\mathbf{p}_{i}^{a}\right)\right)^{\beta_{b}^{\toa}}\label{eq:bias_variance_TOA}
\end{equation}
is proposed; here $\beta_{b}^{\toa}$ is a dimensionless parameter
whereas $\sigma_{b,0}^{\toa}$ is expressed in meters. For RSS systems,
instead, our experimental results have evidenced that the bias variance
is mostly constant or even decreases as the number of obstructions
$N_{o}\left(\mathbf{p},\mathbf{p}_{i}^{a}\right)$ gets larger and
that the linear model
\begin{equation}
\sigma_{b,i}^{\rss}(\mathbf{p})=\max\left\{ \sigma_{b,0}^{\rss}-\sigma_{b,m}^{\rss}\cdot N_{o}\left(\mathbf{p},\mathbf{p}_{i}^{a}\right),0\right\} \label{eq:bias_variance_RSS}
\end{equation}
can be adopted, where both the parameters $\sigma_{b,0}^{\rss}$ and
$\sigma_{b,m}^{\rss}$ are expressed in meters.

\emph{Noise model fit} - In various technical papers (e.g., see \cite{Yu2009})
the additive noise term $n_{i}(\mathbf{p})$ is assumed to have the
same variance for all LOS (NLOS) links, regardless of other factors,
like the link distance. This noise model accounts for the inaccuracy
of the algorithm adopted in extrapolating the observations $\left\{ z_{i}\right\} $
in NLOS conditions, but does not account for the dependence of the
performance of such an algorithm on the \ac{SNR}. In this work,
based on experimental evidence, we have decided to employ, for both
TOA, TDOA and RSS systems, the model adopted in \cite{Venkatesh2007,Jourdan2006,Venkatesh2007a}
and expressed by 
\begin{equation}
\sigma_{n,i}(\mathbf{p})=\sigma_{n,0}\left(d_{i}(\mathbf{p})/d_{0}\right)^{\beta_{n}}\label{eq:noise_model}
\end{equation}
where $\sigma_{n,0}^{2}$ is the noise variance characterizing the
transmitter-receiver distance $d_{0}$ and $\beta_{n}$ is the noise
path loss exponent.

Finally, substituting (\ref{eq:bias_mean_TOA}), (\ref{eq:bias_variance_TOA})
and (\ref{eq:noise_model}) (or (\ref{eq:bias_mean_RSS}), (\ref{eq:bias_variance_RSS})
and (\ref{eq:noise_model})) in (\ref{eq:likelihood_with_maps}) produces
(\ref{eq:likelihood_with_maps_toa_final}) (or (\ref{eq:likelihood_with_maps_rss_final})).

\subsection{Fitting Measurements in the Map-Unaware Case\label{app:nomap_fitting}}

A simpler procedure can be adopted for extracting regression fits
in the map-unaware case. Such a procedure consists of the three steps
described below.

\emph{Data modelling} - On the basis of (\ref{eq:likelihood_no_maps_gaussian}),
each of the residuals $\left\{ r_{(m,i)}\right\} $ follows a Gaussian
distribution, i.e.,
\begin{align}
f\left(\left\{ r_{(m,i)}\right\} \right)= & \prod_{i\in\mathcal{Z}^{\los}}\mathcal{N}\left(r_{(m,i)};0,\sigma_{n,i}^{2}(\tilde{\mathbf{p}})\right)\nonumber \\
 & \prod_{i\in\mathcal{Z}^{\nlos}}\mathcal{N}\left(r_{(m,i)};\kappa_{b},\sigma_{n,i}^{2}(\tilde{\mathbf{p}})+\gamma_{b}^{2}\right)\label{eq:pdf_single_measurement_nomap_gaussian}
\end{align}
or a mixed Gaussian-exponential distribution (\ref{eq:likelihood_no_maps_exp}),
i.e., 
\begin{align}
f\left(\left\{ r_{(m,i)}\right\} \right)= & \prod_{i\in\mathcal{Z}^{\los}}\mathcal{N}\left(r_{(m,i)};0,\sigma_{n,i}^{2}(\tilde{\mathbf{p}})\right)\nonumber \\
 & \prod_{i\in\mathcal{Z}^{\nlos}}\mathcal{E}\left(z_{i};0,\nu_{b}\right)\label{eq:pdf_single_measurement_nomap_exp}
\end{align}
where $\mathcal{Z}^{\los}$ and $\mathcal{Z}^{\nlos}$ are known in
this fitting phase.

\emph{Bias model fit}\textbf{ - }On the basis of experimental data
the model (\ref{eq:pdf_single_measurement_nomap_gaussian}) has been
adopted for modelling the RSS measurements acquired by means of a
couple of radio devices operating at $169\mhz$. The model (\ref{eq:pdf_single_measurement_nomap_exp}),
instead, has been employed to describe the TOA data collected in the
Newcom++ database. The \acp{MLE} of the parameters $\kappa_{b}$
and $\gamma_{b}$ appearing in the model (\ref{eq:pdf_single_measurement_nomap_gaussian})
(see Sec. \ref{sec:statistical_model}) are easily shown (using a
procedure similar to that adopted in the bias model fitting for the
map-aware case) to coincide with the sample average and the sample
variance, respectively, of the residuals $\left\{ r_{(m,i)},\forall(m,i)\in\mathcal{I}_{p}\right\} $,
where $\mathcal{I}_{p}\triangleq\left\{ (m,i)|t_{o,(m,i)}>0\right\} $.\textbf{
}Similarly, if the model (\ref{eq:pdf_single_measurement_nomap_exp})
is used, the \ac{MLE} of $\nu_{b}$ coincides with the sample average
of the residuals $\left\{ r_{(m,i)},\forall(m,i)\in\mathcal{I}_{p}\right\} $.

\emph{Noise model fit} - Following the approach adopted in the map-aware
case, the noise is assumed to depend on the link distance and the
ML estimate $\hat{\sigma}_{n}(t)$ of the noise standard deviation
is evaluated for those links identified by a couple of indices $(m,i)$
such that $t_{d,(m,i)}=t$. Unluckily, no closed-form solution is
available for this estimate, which can be expressed as (note the similarity
with (\ref{eq:noise_var_estimator_final}))
\begin{multline}
\hat{\sigma}_{n}(t)=\arg\min_{\tilde{\sigma}}\sum_{(m,i)\in\mathcal{I}_{n}(t)}\left\{ \frac{\left(r{}_{(m,i)}-\mu_{b,(m,i)}\right)^{2}}{\sigma_{b,(m,i)}^{2}+\tilde{\sigma}^{2}}\right.\\
\left.+2\ln\sqrt{\sigma_{b,(m,i)}^{2}+\tilde{\sigma}^{2}}\vphantom{\frac{\left(r{}_{(m,i)}-\mu_{b,(m,i)}\right)^{2}}{\sigma_{b,(m,i)}^{2}+\tilde{\sigma}^{2}}}\right\} \label{eq:noise_var_estimator_final_nomap}
\end{multline}
for $t=1,...,N_{d,\mymax}$. It is also important to mention that
$\mu_{b,(m,i)}$ and $\sigma_{b,(m,i)}$ are defined as: a) $\mu_{b,(m,i)}\triangleq\kappa_{b}$
and $\sigma_{b,(m,i)}\triangleq\gamma_{b}$, respectively, if the
NLOS bias is modelled as a Gaussian rv (see (\ref{eq:likelihood_no_maps_gaussian})
and (\ref{eq:pdf_single_measurement_nomap_gaussian})); b) $\mu_{b,(m,i)}\triangleq\nu_{b}$
and $\sigma_{b,(m,i)}\triangleq\nu_{b}^{2}$, respectively, if the
NLOS bias is modelled as an exponential random variable (see (\ref{eq:likelihood_no_maps_exp})
and (\ref{eq:pdf_single_measurement_nomap_exp})). In both cases we
have found that the quantities $\left\{ \hat{\sigma}_{n}(t)\right\} $
do exhibit a monotonic dependence on the distance (due to all propagation
mechanisms not accounted for by a map-unaware model); despite this,
the model (\ref{eq:noise_model}) has been selected as a regression
function for $\sigma_{n,i}(\mathbf{p})$.

Finally, substituting (\ref{eq:noise_model}) in (\ref{eq:likelihood_no_maps_gaussian})
produces the likelihood function (\ref{eq:likelihood_no_maps_rss_final_gaussian})
proposed for map-unaware RSS systems. Similarly,\emph{ }substituting
(\ref{eq:noise_model}) in (\ref{eq:likelihood_no_maps_exp}) yields
the likelihood function (\ref{eq:likelihood_no_maps_toa_final_exp})
referring to map-unaware TOA systems.

\bibliographystyle{IEEEtran}

\begin{IEEEbiography}[{\includegraphics[width=1in,height=1.25in,clip,keepaspectratio]{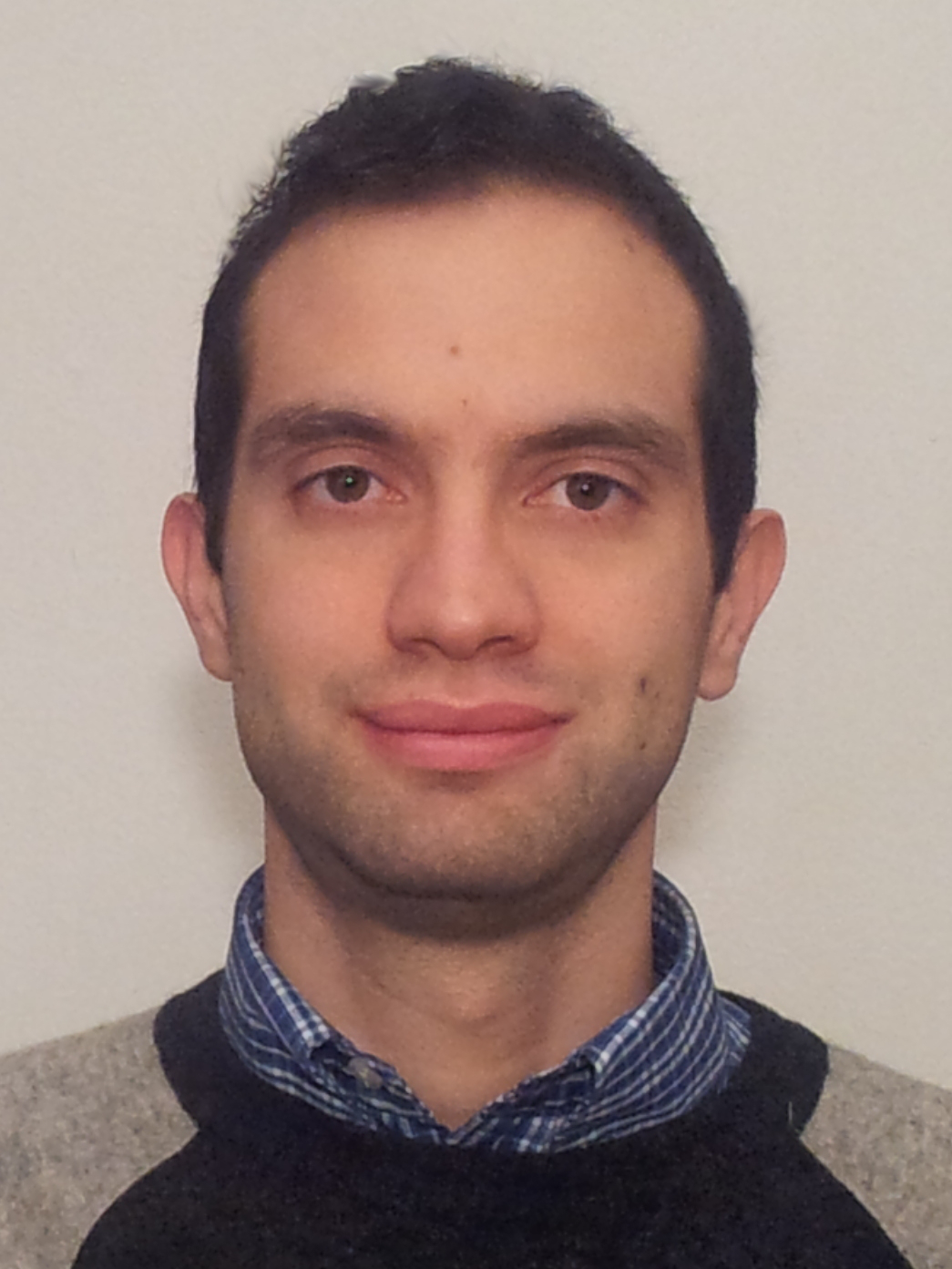}}]{Francesco Montorsi}
(S’06) received both the Laurea degree (\emph{cum laude}) and the Laurea Specialistica degree (\emph{cum laude}) in Electronic Engineering from the University of Modena and Reggio Emilia, Italy, in 2007 and 2009, respectively. He received the Ph.D degree in Information and Communications Technologies (ICT) from the University of Modena and Reggio Emilia in 2013.

He is employed as embedded system engineer for an ICT company since 2013. In 2011 he was a visiting PhD student at the Wireless Communications and Network Science Laboratory of Massachusetts Institute of Technology (MIT). His research interests are in the area of localization and navigation systems, with emphasis on statistical signal processing (linear and non-linear filtering, detection and estimation problems), model-based design and model-based performance assessment.

Dr. Montorsi is a member of IEEE Communications Society and served as a reviewer for the \textsc{IEEE Transactions on Wireless Communications},\textsc{IEEE Transactions on Signal Processing}, \textsc{IEEE Wireless Communications Letters} and several IEEE conferences.
He received the GTTI Award for PhD Theses in the field of Communication Technologies in 2013 from the Italian Telecommunications and Information Theory Group (GTTI).
\end{IEEEbiography}

\vspace*{-0.5in}

\begin{IEEEbiography}[{\includegraphics[width=1in,height=1.25in,clip,keepaspectratio]{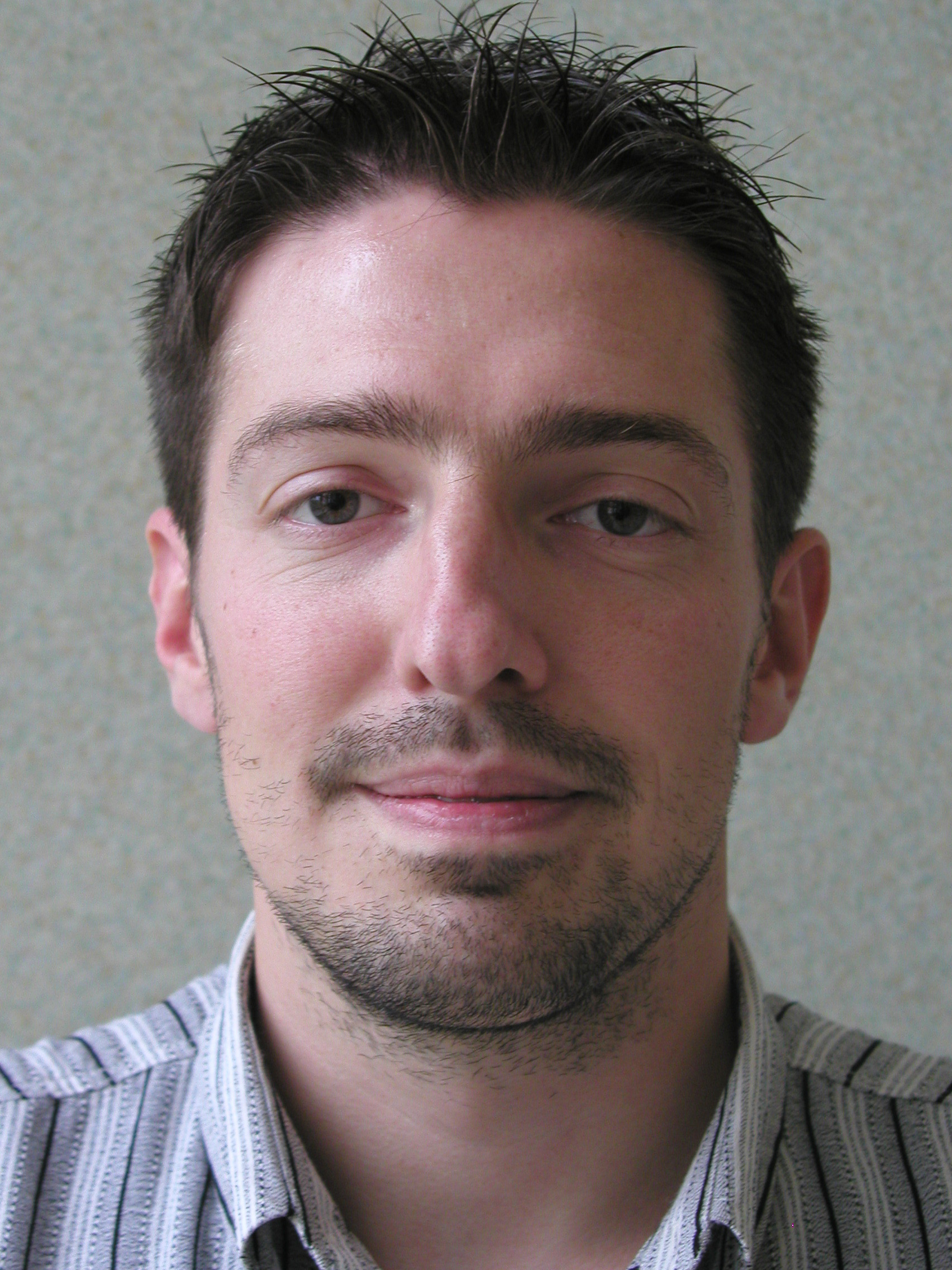}}]{Fabrizio Pancaldi}
was born in Modena, Italy, in July 1978. He received the Dr. Eng. Degree in Electronic Engineering (cum laude) and the Ph. D. degree in 2006, both from the University of Modena and Reggio Emilia, Italy. From March 2006 he is holding the position of Assistant Professor at the same university and he gives the courses of Telecommunication Networks and ICT Systems. He works in the field of digital communications, both radio and powerline. His particular interests lie in the wide area of digital communications, with emphasis on channel equalization, statistical channel modelling, space-time coding, radio localization, channel estimation and clock synchronization.
\end{IEEEbiography}

\vspace*{-0.5in}

\begin{IEEEbiography}[{\includegraphics[width=1in,height=1.25in,clip,keepaspectratio]{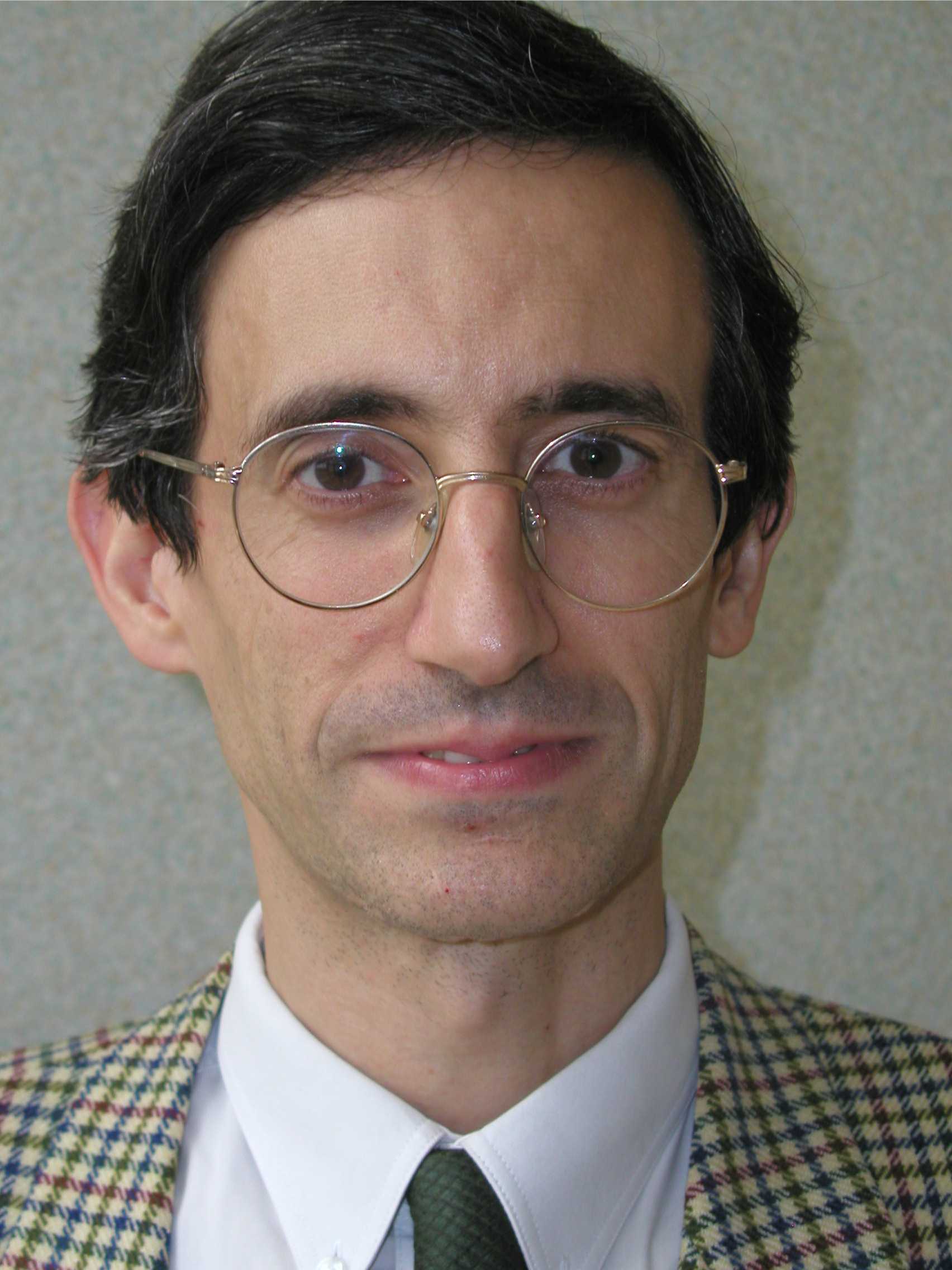}}]{Giorgio M. Vitetta}
(S’89-M’91-SM’99) was born in Reggio Calabria, Italy, in April 1966. He received the Dr. Ing. degree in electronic engineering (\emph{cum laude}) in 1990, and the Ph.D. degree in 1994, both from the University of Pisa, Pisa, Italy. From 1995 to 1998, he was a Research Fellow with the Department of Information Engineering, University of Pisa. From 1998 to 2001, he was an Associate Professor with the University of Modena and Reggio Emilia, Modena, Italy, where he is currently a full Professor. 

His main research interests lie in the broad area of communication theory, with particular emphasis on detection/equalization/synchronization algorithms for wireless communications, statistical modelling of wireless and powerline channels, ultrawideband communication techniques and applications of game theory to wireless communications. 

Dr. Vitetta is serving as an Editor of the \textsc{IEEE Wireless Communications Letters} and as an Area Editor of the \textsc{IEEE Transactions on Communications}.
\end{IEEEbiography}

\end{document}

%% file: fig1.pdf_tex
\begingroup%
  \makeatletter%
  \providecommand\color[2][]{%
    \errmessage{(Inkscape) Color is used for the text in Inkscape, but the package 'color.sty' is not loaded}%
    \renewcommand\color[2][]{}%
  }%
  \providecommand\transparent[1]{%
    \errmessage{(Inkscape) Transparency is used (non-zero) for the text in Inkscape, but the package 'transparent.sty' is not loaded}%
    \renewcommand\transparent[1]{}%
  }%
  \providecommand\rotatebox[2]{#2}%
  \ifx\svgwidth\undefined%
    \setlength{\unitlength}{574.42758789bp}%
    \ifx\svgscale\undefined%
      \relax%
    \else%
      \setlength{\unitlength}{\unitlength * \real{\svgscale}}%
    \fi%
  \else%
    \setlength{\unitlength}{\svgwidth}%
  \fi%
  \global\let\svgwidth\undefined%
  \global\let\svgscale\undefined%
  \makeatother%
  \begin{picture}(1,0.46933675)%
    \put(0,0){\includegraphics[width=\unitlength]{fig1.pdf}}%
    \put(0.15130533,0.40489819){\color[rgb]{0,0,0}\makebox(0,0)[lb]{\smash{$\mathbf{p}_{1}^{a}$}}}%
    \put(0.18036781,0.04317065){\color[rgb]{0,0,0}\makebox(0,0)[lb]{\smash{$\mathbf{p}_{2}^{a}$}}}%
    \put(0.58632284,0.35085882){\color[rgb]{0,0,0}\makebox(0,0)[lb]{\smash{$\mathbf{p}_{6}^{a}$}}}%
    \put(0.56865291,0.07552785){\color[rgb]{0,0,0}\makebox(0,0)[lb]{\smash{$\mathbf{p}_{3}^{a}$}}}%
    \put(0.29008343,0.15092602){\color[rgb]{0,0,0}\makebox(0,0)[lb]{\smash{$\mathbf{p}$}}}%
    \put(0.17068149,0.25884289){\color[rgb]{0,0,0}\makebox(0,0)[lb]{\smash{$z_1$}}}%
    \put(0.05329755,0.14344847){\color[rgb]{0,0,0}\makebox(0,0)[lb]{\smash{$\mathcal{R}$}}}%
    \put(0.7812995,0.35085882){\color[rgb]{0,0,0}\makebox(0,0)[lb]{\smash{$\mathbf{p}_{5}^{a}$}}}%
    \put(0.72377872,0.15936489){\color[rgb]{0,0,0}\makebox(0,0)[lb]{\smash{$\mathbf{p}_{4}^{a}$}}}%
    \put(0.92452904,0.03680821){\color[rgb]{0,0,0}\makebox(0,0)[lb]{\smash{$\mathcal{B}\left(\mathcal{R}\right)$}}}%
  \end{picture}%
\endgroup%

%% file: fig2.pdf_tex
\begingroup%
  \makeatletter%
  \providecommand\color[2][]{%
    \errmessage{(Inkscape) Color is used for the text in Inkscape, but the package 'color.sty' is not loaded}%
    \renewcommand\color[2][]{}%
  }%
  \providecommand\transparent[1]{%
    \errmessage{(Inkscape) Transparency is used (non-zero) for the text in Inkscape, but the package 'transparent.sty' is not loaded}%
    \renewcommand\transparent[1]{}%
  }%
  \providecommand\rotatebox[2]{#2}%
  \ifx\svgwidth\undefined%
    \setlength{\unitlength}{866.73544922bp}%
    \ifx\svgscale\undefined%
      \relax%
    \else%
      \setlength{\unitlength}{\unitlength * \real{\svgscale}}%
    \fi%
  \else%
    \setlength{\unitlength}{\svgwidth}%
  \fi%
  \global\let\svgwidth\undefined%
  \global\let\svgscale\undefined%
  \makeatother%
  \begin{picture}(1,0.72925357)%
    \put(0,0){\includegraphics[width=\unitlength]{fig2.pdf}}%
    \put(0.84774028,0.16495975){\color[rgb]{0,0,0}\makebox(0,0)[lb]{\smash{$\mathbf{p}_{12}$}}}%
    \put(0.71819054,0.12354672){\color[rgb]{0,0,0}\makebox(0,0)[lb]{\smash{$\mathbf{p}_{14}$}}}%
    \put(0.0736245,0.65125883){\color[rgb]{0,0,0}\makebox(0,0)[lb]{\smash{$\mathbf{p}_1$}}}%
    \put(0.26968647,0.63843242){\color[rgb]{0,0,0}\makebox(0,0)[lb]{\smash{$\mathbf{p}_2$}}}%
    \put(0.50247736,0.63245678){\color[rgb]{0,0,0}\makebox(0,0)[lb]{\smash{$\mathbf{p}_3$}}}%
    \put(0.10863409,0.53863312){\color[rgb]{0,0,0}\makebox(0,0)[lb]{\smash{$\mathbf{p}_4$}}}%
    \put(0.0834165,0.43576832){\color[rgb]{0,0,0}\makebox(0,0)[lb]{\smash{$\mathbf{p}_5$}}}%
    \put(0.12796935,0.3430398){\color[rgb]{0,0,0}\makebox(0,0)[lb]{\smash{$\mathbf{p}_8$}}}%
    \put(0.70825974,0.33809722){\color[rgb]{0,0,0}\makebox(0,0)[lb]{\smash{$\mathbf{p}_6$}}}%
    \put(0.12083402,0.24320944){\color[rgb]{0,0,0}\makebox(0,0)[lb]{\smash{$\mathbf{p}_{10}$}}}%
    \put(0.40533714,0.18452857){\color[rgb]{0,0,0}\makebox(0,0)[lb]{\smash{$\mathbf{p}_{11}$}}}%
    \put(0.56539501,0.09010102){\color[rgb]{0,0,0}\makebox(0,0)[lb]{\smash{$\mathbf{p}_{16}$}}}%
    \put(0.1206777,0.13330536){\color[rgb]{0,0,0}\makebox(0,0)[lb]{\smash{$\mathbf{p}_{13}$}}}%
    \put(0.98290742,0.48893609){\color[rgb]{0,0,0}\makebox(0,0)[lb]{\smash{$x$}}}%
    \put(0.76047035,0.71830364){\color[rgb]{0,0,0}\makebox(0,0)[lb]{\smash{$y$}}}%
    \put(0.29434389,0.30166274){\color[rgb]{0,0,0}\makebox(0,0)[lb]{\smash{$\mathbf{p}_7$}}}%
    \put(0.44651435,0.32479456){\color[rgb]{0,0,0}\makebox(0,0)[lb]{\smash{$\mathbf{p}_9$}}}%
    \put(0.27891868,0.09744415){\color[rgb]{0,0,0}\makebox(0,0)[lb]{\smash{$\mathbf{p}_{15}$}}}%
    \put(0.01010166,0.05976623){\color[rgb]{0,0,0}\rotatebox{90}{\makebox(0,0)[lb]{\smash{$5\meter$}}}}%
  \end{picture}%
\endgroup%

%% file: fig4.tikz
%
%
%
%
\begin{tikzpicture}

\begin{axis}[%
width=\figurewidth,
height=\figureheight,
scale only axis,
xmin=0, xmax=55,
xlabel={Distance $d$ [m]},
ymin=-80, ymax=-20,
ytick={-80,-70,...,-20},
ylabel={RSS $P_{rx}$ [dBm]},
axis on top,
legend style={nodes=right,draw=none,font=\footnotesize}]

\addplot [
color=black,
only marks,
mark=+,
mark options={solid}
]
plot [error bars/.cd, y dir = both, y explicit]
coordinates{
 (1.18258192105241,-36.16)+-(0.0,0.650274667242343)(2.37118114027588,-39.84)+-(0.0,1.05675668622325)(3.56742203839131,-34.96)+-(0.0,0.197948663722158)(4.76555348306994,-35.74)+-(0.0,0.486973158544551)(5.96443626841632,-39)+-(0.0,0)(7.16369318159286,-42.12)+-(0.0,0.520596204522533)(8.36316327713384,-44.32)+-(0.0,0.47121207149916)(9.56276633616026,-44.54)+-(0.0,0.503457433905886)(10.7624578977109,-44)+-(0.0,0)(11.9622113340302,-43.28)+-(0.0,0.640153042925945)(13.1620097249622,-44.28)+-(0.0,0.453557367611076)(14.3618418038913,-47.92)+-(0.0,0.665168384389872)(15.5616997786232,-52.4)+-(0.0,0.534522483824851)(16.7615780879964,-50.32)+-(0.0,1.40610911992734)(17.9614726567729,-50.28)+-(0.0,0.701019665507729)(19.1613804304387,-53.26)+-(0.0,0.443087497693455)(20.3612990744697,-55.3)+-(0.0,0.462910049886275)(21.5612267740034,-57.66)+-(0.0,0.847806292854727)(22.7611620968702,-56.12)+-(0.0,0.328260722659316)(23.9611038977757,-54.44)+-(0.0,0.577114567906406)(25.1610512498981,-56)+-(0.0,0)(26.3610033951669,-57.38)+-(0.0,0.635353027369632)(27.5609597075283,-59.14)+-(0.0,0.350509832753865)(28.7609196654071,-59.3)+-(0.0,0.543983793275991)(29.9608828307845,-59.46)+-(0.0,0.503457433905886)(31.1608488331111,-60.38)+-(0.0,0.696639162299235)(32.3608173567974,-60.68)+-(0.0,2.26274169979695)(33.5607881313893,-59.98)+-(0.0,0.141421356237309)(34.7607609237773,-62.12)+-(0.0,0.68927646181799)(35.960735531966,-62.94)+-(0.0,0.313635691430001)(37.1607117800507,-65.62)+-(0.0,0.49031435147802)(38.3606895141367,-65.1)+-(0.0,0.30304576336566)(39.5606685990012,-66)+-(0.0,0)(40.7606489153448,-66.44)+-(0.0,0.732900305050322)(41.9606303575149,-67.56)+-(0.0,0.5014265364224)(43.1606128316084,-69.1)+-(0.0,0.36421567954234)(44.3605962538828,-70)+-(0.0,0.404061017820879)(45.5605805494179,-69.02)+-(0.0,0.141421356237316)(46.7605656509842,-72.04)+-(0.0,0.197948663722158)(47.9605514980802,-71.7)+-(0.0,0.462910049886275)(49.1605380361119,-76.52)+-(0.0,0.543608502584007)(50.3605252156885,-77.72)+-(0.0,0.496518491342371)(51.5605129920174,-77.88)+-(0.0,0.38544964466378)(52.7605013243809,-78.14)+-(0.0,0.350509832753872) 
};

\addlegendentry{Measured data};

\addplot [
color=black,
dotted,
line width=2.0pt
]
coordinates{
 (1.18258192105241,-17.0883893320018)(1.28258192105241,-18.2121651431147)(1.38258192105241,-19.2515277772545)(1.48258192105241,-20.2182775069461)(1.58258192105241,-21.1219010618048)(1.68258192105241,-21.9701394919437)(1.78258192105241,-22.7693918301477)(1.88258192105241,-23.5250084270288)(1.98258192105241,-24.2415082761901)(2.08258192105241,-24.9227428045364)(2.18258192105241,-25.5720212124666)(2.28258192105241,-26.1922077114297)(2.38258192105241,-26.785797896545)(2.48258192105241,-27.3549794064924)(2.58258192105241,-27.9016805970854)(2.68258192105241,-28.4276099629794)(2.78258192105241,-28.9342883407662)(2.88258192105241,-29.4230754237594)(2.98258192105241,-29.8951917531697)(3.08258192105241,-30.3517370813074)(3.18258192105241,-30.7937058021537)(3.28258192105241,-31.2219999939654)(3.38258192105241,-31.6374405040915)(3.48258192105241,-32.0407764184025)(3.58258192105241,-32.4326931898405)(3.68258192105241,-32.8138196476744)(3.78258192105241,-33.1847340674641)(3.88258192105241,-33.5459694488462)(3.98258192105241,-33.8980181220542)(4.08258192105241,-34.2413357830884)(4.18258192105241,-34.5763450405194)(4.28258192105241,-34.9034385431787)(4.38258192105241,-35.2229817467975)(4.48258192105241,-35.5353153684807)(4.58258192105241,-35.8407575703527)(4.68258192105241,-36.1396059074614)(4.78258192105241,-36.4321390698366)(4.88258192105241,-36.7186184442663)(4.98258192105241,-36.9992895177267)(5.08258192105241,-37.2743831413491)(5.18258192105241,-37.5441166712297)(5.28258192105241,-37.8086950002092)(5.38258192105241,-38.0683114928909)(5.48258192105241,-38.3231488345872)(5.58258192105241,-38.5733798035306)(5.68258192105241,-38.8191679745245)(5.78258192105241,-39.0606683612118)(5.88258192105241,-39.2980280032759)(5.98258192105241,-39.531386504145)(6.08258192105241,-39.7608765241232)(6.18258192105241,-39.9866242333108)(6.28258192105241,-40.2087497281836)(6.38258192105241,-40.4273674152776)(6.48258192105241,-40.6425863650456)(6.58258192105241,-40.8545106386264)(6.68258192105241,-41.0632395899781)(6.78258192105241,-41.2688681455701)(6.88258192105241,-41.4714870636054)(6.98258192105241,-41.6711831745456)(7.08258192105241,-41.868039604532)(7.18258192105241,-42.0621359831446)(7.28258192105241,-42.2535486367958)(7.38258192105241,-42.4423507689363)(7.48258192105241,-42.6286126281368)(7.58258192105241,-42.8124016650115)(7.68258192105241,-42.9937826788615)(7.78258192105241,-43.1728179548348)(7.88258192105241,-43.3495673923295)(7.98258192105241,-43.5240886253036)(8.08258192105241,-43.696437135095)(8.18258192105241,-43.8666663563045)(8.28258192105241,-44.0348277762479)(8.38258192105241,-44.20097102844)(8.48258192105241,-44.3651439805346)(8.58258192105241,-44.5273928171118)(8.68258192105241,-44.6877621176694)(8.78258192105241,-44.8462949301484)(8.88258192105241,-45.0030328402964)(8.98258192105241,-45.1580160371479)(9.08258192105241,-45.3112833748795)(9.18258192105241,-45.462872431279)(9.28258192105241,-45.6128195630475)(9.38258192105241,-45.761159958138)(9.48258192105241,-45.9079276853203)(9.58258192105241,-46.0531557411458)(9.68258192105241,-46.1968760944736)(9.78258192105241,-46.3391197287103)(9.88258192105241,-46.4799166819004)(9.98258192105241,-46.6192960847996)(10.0825819210524,-46.7572861970497)(10.1825819210524,-46.8939144415688)(10.2825819210524,-47.0292074372606)(10.3825819210524,-47.1631910301416)(10.4825819210524,-47.295890322975)(10.5825819210524,-47.4273297034995)(10.6825819210524,-47.5575328713291)(10.7825819210524,-47.6865228636004)(10.8825819210524,-47.8143220794358)(10.9825819210524,-47.9409523032883)(11.0825819210524,-48.0664347272273)(11.1825819210524,-48.1907899722244)(11.2825819210524,-48.3140381084912)(11.3825819210524,-48.4361986749198)(11.4825819210524,-48.557290697673)(11.5825819210524,-48.6773327079686)(11.6825819210524,-48.7963427590992)(11.7825819210524,-48.9143384427266)(11.8825819210524,-49.031336904488)(11.9825819210524,-49.1473548589472)(12.0825819210524,-49.2624086039261)(12.1825819210524,-49.3765140342441)(12.2825819210524,-49.4896866548965)(12.3825819210524,-49.6019415936992)(12.4825819210524,-49.7132936134231)(12.5825819210524,-49.8237571234466)(12.6825819210524,-49.9333461909452)(12.7825819210524,-50.0420745516423)(12.8825819210524,-50.149955620141)(12.9825819210524,-50.2570024998567)(13.0825819210524,-50.3632279925682)(13.1825819210524,-50.4686446076055)(13.2825819210524,-50.5732645706901)(13.3825819210524,-50.6770998324443)(13.4825819210524,-50.7801620765828)(13.5825819210524,-50.8824627278028)(13.6825819210524,-50.9840129593837)(13.7825819210524,-51.0848237005108)(13.8825819210524,-51.1849056433339)(13.9825819210524,-51.2842692497725)(14.0825819210524,-51.382924758079)(14.1825819210524,-51.480882189169)(14.2825819210524,-51.5781513527299)(14.3825819210524,-51.6747418531158)(14.4825819210524,-51.7706630950387)(14.5825819210524,-51.8659242890635)(14.6825819210524,-51.9605344569151)(14.7825819210524,-52.0545024366062)(14.8825819210524,-52.1478368873914)(14.9825819210524,-52.2405462945554)(15.0825819210524,-52.3326389740428)(15.1825819210524,-52.4241230769339)(15.2825819210524,-52.5150065937741)(15.3825819210524,-52.6052973587625)(15.4825819210524,-52.6950030538036)(15.5825819210524,-52.7841312124299)(15.6825819210524,-52.8726892235981)(15.7825819210524,-52.9606843353654)(15.8825819210524,-53.0481236584487)(15.9825819210524,-53.1350141696732)(16.0825819210524,-53.2213627153127)(16.1825819210524,-53.3071760143268)(16.2825819210524,-53.3924606614979)(16.3825819210524,-53.4772231304728)(16.4825819210524,-53.5614697767111)(16.5825819210524,-53.6452068403451)(16.6825819210524,-53.728440448953)(16.7825819210524,-53.8111766202499)(16.8825819210524,-53.8934212646983)(16.9825819210524,-53.975180188042)(17.0825819210524,-54.0564590937653)(17.1825819210524,-54.1372635854804)(17.2825819210524,-54.2175991692459)(17.3825819210524,-54.297471255818)(17.4825819210524,-54.3768851628373)(17.5825819210524,-54.4558461169533)(17.6825819210524,-54.5343592558886)(17.7825819210524,-54.6124296304445)(17.8825819210524,-54.6900622064516)(17.9825819210524,-54.7672618666642)(18.0825819210524,-54.8440334126038)(18.1825819210524,-54.9203815663506)(18.2825819210524,-54.9963109722871)(18.3825819210524,-55.0718261987924)(18.4825819210524,-55.1469317398923)(18.5825819210524,-55.2216320168641)(18.6825819210524,-55.295931379798)(18.7825819210524,-55.3698341091173)(18.8825819210524,-55.4433444170584)(18.9825819210524,-55.516466449111)(19.0825819210524,-55.5892042854214)(19.1825819210524,-55.6615619421582)(19.2825819210524,-55.7335433728433)(19.3825819210524,-55.8051524696476)(19.4825819210524,-55.8763930646544)(19.5825819210524,-55.9472689310897)(19.6825819210524,-56.0177837845213)(19.7825819210524,-56.0879412840278)(19.8825819210524,-56.1577450333376)(19.9825819210524,-56.2271985819402)(20.0825819210524,-56.2963054261686)(20.1825819210524,-56.3650690102559)(20.2825819210524,-56.4334927273652)(20.3825819210524,-56.5015799205942)(20.4825819210524,-56.5693338839555)(20.5825819210524,-56.6367578633326)(20.6825819210524,-56.7038550574133)(20.7825819210524,-56.7706286186)(20.8825819210524,-56.8370816538985)(20.9825819210524,-56.9032172257856)(21.0825819210524,-56.9690383530557)(21.1825819210524,-57.034548011648)(21.2825819210524,-57.0997491354535)(21.3825819210524,-57.1646446171037)(21.4825819210524,-57.2292373087406)(21.5825819210524,-57.2935300227687)(21.6825819210524,-57.3575255325898)(21.7825819210524,-57.4212265733211)(21.8825819210524,-57.4846358424965)(21.9825819210524,-57.5477560007518)(22.0825819210524,-57.6105896724951)(22.1825819210524,-57.673139446561)(22.2825819210524,-57.735407876851)(22.3825819210524,-57.7973974829591)(22.4825819210524,-57.8591107507835)(22.5825819210524,-57.9205501331247)(22.6825819210524,-57.9817180502708)(22.7825819210524,-58.042616890569)(22.8825819210524,-58.1032490109858)(22.9825819210524,-58.163616737654)(23.0825819210524,-58.2237223664083)(23.1825819210524,-58.2835681633091)(23.2825819210524,-58.3431563651553)(23.3825819210524,-58.4024891799857)(23.4825819210524,-58.4615687875702)(23.5825819210524,-58.5203973398904)(23.6825819210524,-58.5789769616093)(23.7825819210524,-58.6373097505327)(23.8825819210524,-58.695397778059)(23.9825819210524,-58.7532430896214)(24.0825819210524,-58.8108477051194)(24.1825819210524,-58.8682136193427)(24.2825819210524,-58.9253428023849)(24.3825819210524,-58.9822372000501)(24.4825819210524,-59.0388987342503)(24.5825819210524,-59.0953293033952)(24.6825819210524,-59.1515307827733)(24.7825819210524,-59.2075050249264)(24.8825819210524,-59.2632538600159)(24.9825819210524,-59.3187790961815)(25.0825819210524,-59.3740825198936)(25.1825819210524,-59.4291658962977)(25.2825819210524,-59.484030969553)(25.3825819210524,-59.5386794631631)(25.4825819210524,-59.5931130803015)(25.5825819210524,-59.6473335041297)(25.6825819210524,-59.7013423981096)(25.7825819210524,-59.7551414063099)(25.8825819210524,-59.8087321537061)(25.9825819210524,-59.8621162464752)(26.0825819210524,-59.9152952722845)(26.1825819210524,-59.9682708005748)(26.2825819210524,-60.0210443828387)(26.3825819210524,-60.0736175528926)(26.4825819210524,-60.1259918271449)(26.5825819210524,-60.1781687048579)(26.6825819210524,-60.2301496684058)(26.7825819210524,-60.2819361835267)(26.8825819210524,-60.3335296995715)(26.9825819210524,-60.3849316497465)(27.0825819210524,-60.4361434513527)(27.1825819210524,-60.4871665060203)(27.2825819210524,-60.5380021999385)(27.3825819210524,-60.588651904082)(27.4825819210524,-60.6391169744325)(27.5825819210524,-60.6893987521965)(27.6825819210524,-60.7394985640196)(27.7825819210524,-60.7894177221959)(27.8825819210524,-60.8391575248747)(27.9825819210524,-60.8887192562629)(28.0825819210524,-60.938104186824)(28.1825819210524,-60.9873135734735)(28.2825819210524,-61.0363486597708)(28.3825819210524,-61.0852106761077)(28.4825819210524,-61.133900839894)(28.5825819210524,-61.182420355739)(28.6825819210524,-61.2307704156307)(28.7825819210524,-61.2789521991113)(28.8825819210524,-61.32696687345)(28.9825819210524,-61.3748155938125)(29.0825819210524,-61.422499503428)(29.1825819210524,-61.4700197337529)(29.2825819210524,-61.517377404632)(29.3825819210524,-61.5645736244568)(29.4825819210524,-61.6116094903211)(29.5825819210524,-61.6584860881741)(29.6825819210524,-61.7052044929709)(29.7825819210524,-61.7517657688203)(29.8825819210524,-61.7981709691301)(29.9825819210524,-61.8444211367504)(30.0825819210524,-61.8905173041142)(30.1825819210524,-61.9364604933752)(30.2825819210524,-61.9822517165445)(30.3825819210524,-62.0278919756239)(30.4825819210524,-62.0733822627375)(30.5825819210524,-62.1187235602612)(30.6825819210524,-62.16391684095)(30.7825819210524,-62.2089630680634)(30.8825819210524,-62.253863195488)(30.9825819210524,-62.2986181678593)(31.0825819210524,-62.3432289206808)(31.1825819210524,-62.3876963804411)(31.2825819210524,-62.4320214647297)(31.3825819210524,-62.4762050823502)(31.4825819210524,-62.5202481334325)(31.5825819210524,-62.5641515095426)(31.6825819210524,-62.607916093791)(31.7825819210524,-62.6515427609392)(31.8825819210524,-62.6950323775047)(31.9825819210524,-62.7383858018642)(32.0825819210524,-62.7816038843552)(32.1825819210524,-62.8246874673763)(32.2825819210524,-62.8676373854854)(32.3825819210524,-62.910454465497)(32.4825819210524,-62.9531395265776)(32.5825819210524,-62.9956933803396)(32.6825819210524,-63.0381168309342)(32.7825819210524,-63.080410675142)(32.8825819210524,-63.1225757024636)(32.9825819210524,-63.1646126952075)(33.0825819210524,-63.2065224285771)(33.1825819210524,-63.2483056707568)(33.2825819210524,-63.2899631829965)(33.3825819210524,-63.3314957196945)(33.4825819210524,-63.3729040284797)(33.5825819210524,-63.4141888502923)(33.6825819210524,-63.4553509194633)(33.7825819210524,-63.496390963793)(33.8825819210524,-63.5373097046281)(33.9825819210524,-63.5781078569379)(34.0825819210524,-63.618786129389)(34.1825819210524,-63.6593452244195)(34.2825819210524,-63.6997858383115)(34.3825819210524,-63.7401086612629)(34.4825819210524,-63.7803143774583)(34.5825819210524,-63.8204036651383)(34.6825819210524,-63.8603771966682)(34.7825819210524,-63.9002356386062)(34.8825819210524,-63.9399796517692)(34.9825819210524,-63.9796098912993)(35.0825819210524,-64.0191270067282)(35.1825819210524,-64.0585316420412)(35.2825819210524,-64.0978244357401)(35.3825819210524,-64.1370060209055)(35.4825819210524,-64.1760770252575)(35.5825819210524,-64.2150380712168)(35.6825819210524,-64.2538897759634)(35.7825819210524,-64.2926327514958)(35.8825819210524,-64.3312676046887)(35.9825819210524,-64.3697949373499)(36.0825819210524,-64.4082153462767)(36.1825819210524,-64.4465294233113)(36.2825819210524,-64.4847377553956)(36.3825819210524,-64.5228409246251)(36.4825819210524,-64.560839508302)(36.5825819210524,-64.5987340789877)(36.6825819210524,-64.6365252045548)(36.7825819210524,-64.6742134482379)(36.8825819210524,-64.7117993686842)(36.9825819210524,-64.7492835200028)(37.0825819210524,-64.7866664518142)(37.1825819210524,-64.8239487092983)(37.2825819210524,-64.8611308332423)(37.3825819210524,-64.8982133600875)(37.4825819210524,-64.9351968219762)(37.5825819210524,-64.9720817467971)(37.6825819210524,-65.0088686582306)(37.7825819210524,-65.0455580757936)(37.8825819210524,-65.0821505148834)(37.9825819210524,-65.118646486821)(38.0825819210524,-65.1550464988941)(38.1825819210524,-65.1913510543996)(38.2825819210524,-65.2275606526851)(38.3825819210524,-65.26367578919)(38.4825819210524,-65.2996969554866)(38.5825819210524,-65.3356246393197)(38.6825819210524,-65.3714593246469)(38.7825819210524,-65.4072014916771)(38.8825819210524,-65.4428516169094)(38.9825819210524,-65.478410173171)(39.0825819210524,-65.5138776296554)(39.1825819210524,-65.5492544519587)(39.2825819210524,-65.584541102117)(39.3825819210524,-65.6197380386423)(39.4825819210524,-65.6548457165581)(39.5825819210524,-65.689864587435)(39.6825819210524,-65.7247950994253)(39.7825819210524,-65.7596376972978)(39.8825819210524,-65.7943928224712)(39.9825819210524,-65.8290609130481)(40.0825819210524,-65.863642403848)(40.1825819210524,-65.89813772644)(40.2825819210524,-65.9325473091751)(40.3825819210524,-65.9668715772183)(40.4825819210524,-66.0011109525797)(40.5825819210524,-66.0352658541461)(40.6825819210524,-66.0693366977117)(40.7825819210524,-66.103323896008)(40.8825819210524,-66.1372278587342)(40.9825819210524,-66.171048992587)(41.0825819210524,-66.2047877012893)(41.1825819210524,-66.2384443856197)(41.2825819210524,-66.2720194434407)(41.3825819210524,-66.305513269727)(41.4825819210524,-66.3389262565935)(41.5825819210524,-66.3722587933227)(41.6825819210524,-66.4055112663919)(41.7825819210524,-66.4386840595002)(41.8825819210524,-66.4717775535952)(41.9825819210524,-66.5047921268988)(42.0825819210524,-66.5377281549335)(42.1825819210524,-66.570586010548)(42.2825819210524,-66.6033660639424)(42.3825819210524,-66.6360686826932)(42.4825819210524,-66.6686942317782)(42.5825819210524,-66.701243073601)(42.6825819210524,-66.7337155680147)(42.7825819210524,-66.7661120723463)(42.8825819210524,-66.79843294142)(42.9825819210524,-66.8306785275806)(43.0825819210524,-66.8628491807166)(43.1825819210524,-66.8949452482828)(43.2825819210524,-66.926967075323)(43.3825819210524,-66.9589150044922)(43.4825819210524,-66.9907893760785)(43.5825819210524,-67.0225905280251)(43.6825819210524,-67.0543187959515)(43.7825819210524,-67.0859745131749)(43.8825819210524,-67.1175580107313)(43.9825819210524,-67.1490696173961)(44.0825819210524,-67.1805096597047)(44.1825819210524,-67.2118784619727)(44.2825819210524,-67.2431763463163)(44.3825819210524,-67.2744036326717)(44.4825819210524,-67.3055606388153)(44.5825819210524,-67.3366476803823)(44.6825819210524,-67.3676650708866)(44.7825819210524,-67.3986131217396)(44.8825819210524,-67.4294921422685)(44.9825819210524,-67.4603024397357)(45.0825819210524,-67.4910443193562)(45.1825819210524,-67.5217180843166)(45.2825819210524,-67.5523240357924)(45.3825819210524,-67.582862472966)(45.4825819210524,-67.6133336930444)(45.5825819210524,-67.6437379912763)(45.6825819210524,-67.6740756609693)(45.7825819210524,-67.7043469935069)(45.8825819210524,-67.7345522783656)(45.9825819210524,-67.7646918031309)(46.0825819210524,-67.7947658535143)(46.1825819210524,-67.8247747133692)(46.2825819210524,-67.8547186647072)(46.3825819210524,-67.8845979877136)(46.4825819210524,-67.9144129607639)(46.5825819210524,-67.9441638604386)(46.6825819210524,-67.9738509615387)(46.7825819210524,-68.0034745371015)(46.8825819210524,-68.0330348584151)(46.9825819210524,-68.0625321950334)(47.0825819210524,-68.091966814791)(47.1825819210524,-68.1213389838179)(47.2825819210524,-68.1506489665533)(47.3825819210524,-68.179897025761)(47.4825819210524,-68.2090834225422)(47.5825819210524,-68.2382084163507)(47.6825819210524,-68.2672722650059)(47.7825819210524,-68.2962752247069)(47.8825819210524,-68.325217550046)(47.9825819210524,-68.3540994940219)(48.0825819210524,-68.3829213080533)(48.1825819210524,-68.4116832419918)(48.2825819210524,-68.4403855441349)(48.3825819210524,-68.4690284612389)(48.4825819210524,-68.4976122385318)(48.5825819210524,-68.5261371197256)(48.6825819210524,-68.554603347029)(48.7825819210524,-68.5830111611595)(48.8825819210524,-68.611360801356)(48.9825819210524,-68.6396525053904)(49.0825819210524,-68.6678865095801)(49.1825819210524,-68.6960630487993)(49.2825819210524,-68.724182356491)(49.3825819210524,-68.7522446646789)(49.4825819210524,-68.7802502039781)(49.5825819210524,-68.8081992036074)(49.6825819210524,-68.8360918913999)(49.7825819210524,-68.8639284938142)(49.8825819210524,-68.8917092359459)(49.9825819210524,-68.919434341538)(50.0825819210524,-68.9471040329919)(50.1825819210524,-68.9747185313783)(50.2825819210524,-69.0022780564473)(50.3825819210524,-69.0297828266395)(50.4825819210524,-69.0572330590961)(50.5825819210524,-69.0846289696689)(50.6825819210524,-69.1119707729311)(50.7825819210524,-69.1392586821868)(50.8825819210524,-69.1664929094813)(50.9825819210524,-69.1936736656111)(51.0825819210524,-69.2208011601332)(51.1825819210524,-69.2478756013754)(51.2825819210524,-69.2748971964455)(51.3825819210524,-69.3018661512407)(51.4825819210524,-69.3287826704576)(51.5825819210524,-69.3556469576008)(51.6825819210524,-69.3824592149928)(51.7825819210524,-69.4092196437825)(51.8825819210524,-69.4359284439548)(51.9825819210524,-69.4625858143393)(52.0825819210524,-69.4891919526192)(52.1825819210524,-69.5157470553401)(52.2825819210524,-69.5422513179187)(52.3825819210524,-69.5687049346515)(52.4825819210524,-69.5951080987234)(52.5825819210524,-69.6214610022159)(52.6825819210524,-69.6477638361158) 
};

\addlegendentry{Logarithmic model (\ref{eq:logdistance_path_loss})};

\addplot [
color=black,
solid,
line width=1.5pt
]
coordinates{
 (1.18258192105241,-36.3382767576586)(1.28258192105241,-36.4171970090232)(1.38258192105241,-36.4961172603879)(1.48258192105241,-36.5750375117525)(1.58258192105241,-36.6539577631172)(1.68258192105241,-36.7328780144818)(1.78258192105241,-36.8117982658465)(1.88258192105241,-36.8907185172112)(1.98258192105241,-36.9696387685758)(2.08258192105241,-37.0485590199405)(2.18258192105241,-37.1274792713051)(2.28258192105241,-37.2063995226698)(2.38258192105241,-37.2853197740344)(2.48258192105241,-37.3642400253991)(2.58258192105241,-37.4431602767637)(2.68258192105241,-37.5220805281284)(2.78258192105241,-37.6010007794931)(2.88258192105241,-37.6799210308577)(2.98258192105241,-37.7588412822224)(3.08258192105241,-37.837761533587)(3.18258192105241,-37.9166817849517)(3.28258192105241,-37.9956020363163)(3.38258192105241,-38.074522287681)(3.48258192105241,-38.1534425390457)(3.58258192105241,-38.2323627904103)(3.68258192105241,-38.311283041775)(3.78258192105241,-38.3902032931396)(3.88258192105241,-38.4691235445043)(3.98258192105241,-38.5480437958689)(4.08258192105241,-38.6269640472336)(4.18258192105241,-38.7058842985982)(4.28258192105241,-38.7848045499629)(4.38258192105241,-38.8637248013276)(4.48258192105241,-38.9426450526922)(4.58258192105241,-39.0215653040569)(4.68258192105241,-39.1004855554215)(4.78258192105241,-39.1794058067862)(4.88258192105241,-39.2583260581508)(4.98258192105241,-39.3372463095155)(5.08258192105241,-39.4161665608802)(5.18258192105241,-39.4950868122448)(5.28258192105241,-39.5740070636095)(5.38258192105241,-39.6529273149741)(5.48258192105241,-39.7318475663388)(5.58258192105241,-39.8107678177034)(5.68258192105241,-39.8896880690681)(5.78258192105241,-39.9686083204327)(5.88258192105241,-40.0475285717974)(5.98258192105241,-40.1264488231621)(6.08258192105241,-40.2053690745267)(6.18258192105241,-40.2842893258914)(6.28258192105241,-40.363209577256)(6.38258192105241,-40.4421298286207)(6.48258192105241,-40.5210500799853)(6.58258192105241,-40.59997033135)(6.68258192105241,-40.6788905827146)(6.78258192105241,-40.7578108340793)(6.88258192105241,-40.836731085444)(6.98258192105241,-40.9156513368086)(7.08258192105241,-40.9945715881733)(7.18258192105241,-41.0734918395379)(7.28258192105241,-41.1524120909026)(7.38258192105241,-41.2313323422672)(7.48258192105241,-41.3102525936319)(7.58258192105241,-41.3891728449965)(7.68258192105241,-41.4680930963612)(7.78258192105241,-41.5470133477259)(7.88258192105241,-41.6259335990905)(7.98258192105241,-41.7048538504552)(8.08258192105241,-41.7837741018198)(8.18258192105241,-41.8626943531845)(8.28258192105241,-41.9416146045491)(8.38258192105241,-42.0205348559138)(8.48258192105241,-42.0994551072785)(8.58258192105241,-42.1783753586431)(8.68258192105241,-42.2572956100078)(8.78258192105241,-42.3362158613724)(8.88258192105241,-42.4151361127371)(8.98258192105241,-42.4940563641017)(9.08258192105241,-42.5729766154664)(9.18258192105241,-42.651896866831)(9.28258192105241,-42.7308171181957)(9.38258192105241,-42.8097373695604)(9.48258192105241,-42.888657620925)(9.58258192105241,-42.9675778722897)(9.68258192105241,-43.0464981236543)(9.78258192105241,-43.125418375019)(9.88258192105241,-43.2043386263836)(9.98258192105241,-43.2832588777483)(10.0825819210524,-43.3621791291129)(10.1825819210524,-43.4410993804776)(10.2825819210524,-43.5200196318423)(10.3825819210524,-43.5989398832069)(10.4825819210524,-43.6778601345716)(10.5825819210524,-43.7567803859362)(10.6825819210524,-43.8357006373009)(10.7825819210524,-43.9146208886655)(10.8825819210524,-43.9935411400302)(10.9825819210524,-44.0724613913948)(11.0825819210524,-44.1513816427595)(11.1825819210524,-44.2303018941242)(11.2825819210524,-44.3092221454888)(11.3825819210524,-44.3881423968535)(11.4825819210524,-44.4670626482181)(11.5825819210524,-44.5459828995828)(11.6825819210524,-44.6249031509474)(11.7825819210524,-44.7038234023121)(11.8825819210524,-44.7827436536767)(11.9825819210524,-44.8616639050414)(12.0825819210524,-44.9405841564061)(12.1825819210524,-45.0195044077707)(12.2825819210524,-45.0984246591354)(12.3825819210524,-45.1773449105)(12.4825819210524,-45.2562651618647)(12.5825819210524,-45.3351854132293)(12.6825819210524,-45.414105664594)(12.7825819210524,-45.4930259159587)(12.8825819210524,-45.5719461673233)(12.9825819210524,-45.650866418688)(13.0825819210524,-45.7297866700526)(13.1825819210524,-45.8087069214173)(13.2825819210524,-45.8876271727819)(13.3825819210524,-45.9665474241466)(13.4825819210524,-46.0454676755112)(13.5825819210524,-46.1243879268759)(13.6825819210524,-46.2033081782406)(13.7825819210524,-46.2822284296052)(13.8825819210524,-46.3611486809699)(13.9825819210524,-46.4400689323345)(14.0825819210524,-46.5189891836992)(14.1825819210524,-46.5979094350638)(14.2825819210524,-46.6768296864285)(14.3825819210524,-46.7557499377931)(14.4825819210524,-46.8346701891578)(14.5825819210524,-46.9135904405225)(14.6825819210524,-46.9925106918871)(14.7825819210524,-47.0714309432518)(14.8825819210524,-47.1503511946164)(14.9825819210524,-47.2292714459811)(15.0825819210524,-47.3081916973457)(15.1825819210524,-47.3871119487104)(15.2825819210524,-47.466032200075)(15.3825819210524,-47.5449524514397)(15.4825819210524,-47.6238727028044)(15.5825819210524,-47.702792954169)(15.6825819210524,-47.7817132055337)(15.7825819210524,-47.8606334568983)(15.8825819210524,-47.939553708263)(15.9825819210524,-48.0184739596276)(16.0825819210524,-48.0973942109923)(16.1825819210524,-48.176314462357)(16.2825819210524,-48.2552347137216)(16.3825819210524,-48.3341549650863)(16.4825819210524,-48.4130752164509)(16.5825819210524,-48.4919954678156)(16.6825819210524,-48.5709157191802)(16.7825819210524,-48.6498359705449)(16.8825819210524,-48.7287562219095)(16.9825819210524,-48.8076764732742)(17.0825819210524,-48.8865967246389)(17.1825819210524,-48.9655169760035)(17.2825819210524,-49.0444372273682)(17.3825819210524,-49.1233574787328)(17.4825819210524,-49.2022777300975)(17.5825819210524,-49.2811979814621)(17.6825819210524,-49.3601182328268)(17.7825819210524,-49.4390384841915)(17.8825819210524,-49.5179587355561)(17.9825819210524,-49.5968789869208)(18.0825819210524,-49.6757992382854)(18.1825819210524,-49.7547194896501)(18.2825819210524,-49.8336397410147)(18.3825819210524,-49.9125599923794)(18.4825819210524,-49.991480243744)(18.5825819210524,-50.0704004951087)(18.6825819210524,-50.1493207464734)(18.7825819210524,-50.228240997838)(18.8825819210524,-50.3071612492027)(18.9825819210524,-50.3860815005673)(19.0825819210524,-50.465001751932)(19.1825819210524,-50.5439220032966)(19.2825819210524,-50.6228422546613)(19.3825819210524,-50.7017625060259)(19.4825819210524,-50.7806827573906)(19.5825819210524,-50.8596030087553)(19.6825819210524,-50.9385232601199)(19.7825819210524,-51.0174435114846)(19.8825819210524,-51.0963637628492)(19.9825819210524,-51.1752840142139)(20.0825819210524,-51.2542042655785)(20.1825819210524,-51.3331245169432)(20.2825819210524,-51.4120447683078)(20.3825819210524,-51.4909650196725)(20.4825819210524,-51.5698852710372)(20.5825819210524,-51.6488055224018)(20.6825819210524,-51.7277257737665)(20.7825819210524,-51.8066460251311)(20.8825819210524,-51.8855662764958)(20.9825819210524,-51.9644865278604)(21.0825819210524,-52.0434067792251)(21.1825819210524,-52.1223270305897)(21.2825819210524,-52.2012472819544)(21.3825819210524,-52.2801675333191)(21.4825819210524,-52.3590877846837)(21.5825819210524,-52.4380080360484)(21.6825819210524,-52.516928287413)(21.7825819210524,-52.5958485387777)(21.8825819210524,-52.6747687901423)(21.9825819210524,-52.753689041507)(22.0825819210524,-52.8326092928717)(22.1825819210524,-52.9115295442363)(22.2825819210524,-52.990449795601)(22.3825819210524,-53.0693700469656)(22.4825819210524,-53.1482902983303)(22.5825819210524,-53.2272105496949)(22.6825819210524,-53.3061308010596)(22.7825819210524,-53.3850510524242)(22.8825819210524,-53.4639713037889)(22.9825819210524,-53.5428915551536)(23.0825819210524,-53.6218118065182)(23.1825819210524,-53.7007320578829)(23.2825819210524,-53.7796523092475)(23.3825819210524,-53.8585725606122)(23.4825819210524,-53.9374928119768)(23.5825819210524,-54.0164130633415)(23.6825819210524,-54.0953333147061)(23.7825819210524,-54.1742535660708)(23.8825819210524,-54.2531738174355)(23.9825819210524,-54.3320940688001)(24.0825819210524,-54.4110143201648)(24.1825819210524,-54.4899345715294)(24.2825819210524,-54.5688548228941)(24.3825819210524,-54.6477750742587)(24.4825819210524,-54.7266953256234)(24.5825819210524,-54.805615576988)(24.6825819210524,-54.8845358283527)(24.7825819210524,-54.9634560797174)(24.8825819210524,-55.042376331082)(24.9825819210524,-55.1212965824467)(25.0825819210524,-55.2002168338113)(25.1825819210524,-55.279137085176)(25.2825819210524,-55.3580573365406)(25.3825819210524,-55.4369775879053)(25.4825819210524,-55.51589783927)(25.5825819210524,-55.5948180906346)(25.6825819210524,-55.6737383419993)(25.7825819210524,-55.7526585933639)(25.8825819210524,-55.8315788447286)(25.9825819210524,-55.9104990960932)(26.0825819210524,-55.9894193474579)(26.1825819210524,-56.0683395988225)(26.2825819210524,-56.1472598501872)(26.3825819210524,-56.2261801015519)(26.4825819210524,-56.3051003529165)(26.5825819210524,-56.3840206042812)(26.6825819210524,-56.4629408556458)(26.7825819210524,-56.5418611070105)(26.8825819210524,-56.6207813583751)(26.9825819210524,-56.6997016097398)(27.0825819210524,-56.7786218611044)(27.1825819210524,-56.8575421124691)(27.2825819210524,-56.9364623638338)(27.3825819210524,-57.0153826151984)(27.4825819210524,-57.0943028665631)(27.5825819210524,-57.1732231179277)(27.6825819210524,-57.2521433692924)(27.7825819210524,-57.331063620657)(27.8825819210524,-57.4099838720217)(27.9825819210524,-57.4889041233863)(28.0825819210524,-57.567824374751)(28.1825819210524,-57.6467446261157)(28.2825819210524,-57.7256648774803)(28.3825819210524,-57.804585128845)(28.4825819210524,-57.8835053802096)(28.5825819210524,-57.9624256315743)(28.6825819210524,-58.0413458829389)(28.7825819210524,-58.1202661343036)(28.8825819210524,-58.1991863856682)(28.9825819210524,-58.2781066370329)(29.0825819210524,-58.3570268883976)(29.1825819210524,-58.4359471397622)(29.2825819210524,-58.5148673911269)(29.3825819210524,-58.5937876424915)(29.4825819210524,-58.6727078938562)(29.5825819210524,-58.7516281452208)(29.6825819210524,-58.8305483965855)(29.7825819210524,-58.9094686479501)(29.8825819210524,-58.9883888993148)(29.9825819210524,-59.0673091506795)(30.0825819210524,-59.1462294020441)(30.1825819210524,-59.2251496534088)(30.2825819210524,-59.3040699047734)(30.3825819210524,-59.3829901561381)(30.4825819210524,-59.4619104075027)(30.5825819210524,-59.5408306588674)(30.6825819210524,-59.6197509102321)(30.7825819210524,-59.6986711615967)(30.8825819210524,-59.7775914129614)(30.9825819210524,-59.856511664326)(31.0825819210524,-59.9354319156907)(31.1825819210524,-60.0143521670553)(31.2825819210524,-60.09327241842)(31.3825819210524,-60.1721926697847)(31.4825819210524,-60.2511129211493)(31.5825819210524,-60.330033172514)(31.6825819210524,-60.4089534238786)(31.7825819210524,-60.4878736752433)(31.8825819210524,-60.5667939266079)(31.9825819210524,-60.6457141779726)(32.0825819210524,-60.7246344293372)(32.1825819210524,-60.8035546807019)(32.2825819210524,-60.8824749320665)(32.3825819210524,-60.9613951834312)(32.4825819210524,-61.0403154347959)(32.5825819210524,-61.1192356861605)(32.6825819210524,-61.1981559375252)(32.7825819210524,-61.2770761888898)(32.8825819210524,-61.3559964402545)(32.9825819210524,-61.4349166916191)(33.0825819210524,-61.5138369429838)(33.1825819210524,-61.5927571943485)(33.2825819210524,-61.6716774457131)(33.3825819210524,-61.7505976970778)(33.4825819210524,-61.8295179484424)(33.5825819210524,-61.9084381998071)(33.6825819210524,-61.9873584511717)(33.7825819210524,-62.0662787025364)(33.8825819210524,-62.145198953901)(33.9825819210524,-62.2241192052657)(34.0825819210524,-62.3030394566304)(34.1825819210524,-62.381959707995)(34.2825819210524,-62.4608799593597)(34.3825819210524,-62.5398002107243)(34.4825819210524,-62.618720462089)(34.5825819210524,-62.6976407134536)(34.6825819210524,-62.7765609648183)(34.7825819210524,-62.8554812161829)(34.8825819210524,-62.9344014675476)(34.9825819210524,-63.0133217189123)(35.0825819210524,-63.0922419702769)(35.1825819210524,-63.1711622216416)(35.2825819210524,-63.2500824730062)(35.3825819210524,-63.3290027243709)(35.4825819210524,-63.4079229757355)(35.5825819210524,-63.4868432271002)(35.6825819210524,-63.5657634784648)(35.7825819210524,-63.6446837298295)(35.8825819210524,-63.7236039811942)(35.9825819210524,-63.8025242325588)(36.0825819210524,-63.8814444839235)(36.1825819210524,-63.9603647352881)(36.2825819210524,-64.0392849866528)(36.3825819210524,-64.1182052380174)(36.4825819210524,-64.1971254893821)(36.5825819210524,-64.2760457407468)(36.6825819210524,-64.3549659921114)(36.7825819210524,-64.4338862434761)(36.8825819210524,-64.5128064948407)(36.9825819210524,-64.5917267462054)(37.0825819210524,-64.67064699757)(37.1825819210524,-64.7495672489347)(37.2825819210524,-64.8284875002993)(37.3825819210524,-64.907407751664)(37.4825819210524,-64.9863280030287)(37.5825819210524,-65.0652482543933)(37.6825819210524,-65.144168505758)(37.7825819210524,-65.2230887571226)(37.8825819210524,-65.3020090084873)(37.9825819210524,-65.3809292598519)(38.0825819210524,-65.4598495112166)(38.1825819210524,-65.5387697625812)(38.2825819210524,-65.6176900139459)(38.3825819210524,-65.6966102653106)(38.4825819210524,-65.7755305166752)(38.5825819210524,-65.8544507680399)(38.6825819210524,-65.9333710194045)(38.7825819210524,-66.0122912707692)(38.8825819210524,-66.0912115221338)(38.9825819210524,-66.1701317734985)(39.0825819210524,-66.2490520248631)(39.1825819210524,-66.3279722762278)(39.2825819210524,-66.4068925275925)(39.3825819210524,-66.4858127789571)(39.4825819210524,-66.5647330303218)(39.5825819210524,-66.6436532816864)(39.6825819210524,-66.7225735330511)(39.7825819210524,-66.8014937844157)(39.8825819210524,-66.8804140357804)(39.9825819210524,-66.9593342871451)(40.0825819210524,-67.0382545385097)(40.1825819210524,-67.1171747898744)(40.2825819210524,-67.196095041239)(40.3825819210524,-67.2750152926037)(40.4825819210524,-67.3539355439683)(40.5825819210524,-67.432855795333)(40.6825819210524,-67.5117760466976)(40.7825819210524,-67.5906962980623)(40.8825819210524,-67.669616549427)(40.9825819210524,-67.7485368007916)(41.0825819210524,-67.8274570521563)(41.1825819210524,-67.9063773035209)(41.2825819210524,-67.9852975548856)(41.3825819210524,-68.0642178062502)(41.4825819210524,-68.1431380576149)(41.5825819210524,-68.2220583089795)(41.6825819210524,-68.3009785603442)(41.7825819210524,-68.3798988117089)(41.8825819210524,-68.4588190630735)(41.9825819210524,-68.5377393144382)(42.0825819210524,-68.6166595658028)(42.1825819210524,-68.6955798171675)(42.2825819210524,-68.7745000685321)(42.3825819210524,-68.8534203198968)(42.4825819210524,-68.9323405712614)(42.5825819210524,-69.0112608226261)(42.6825819210524,-69.0901810739908)(42.7825819210524,-69.1691013253554)(42.8825819210524,-69.2480215767201)(42.9825819210524,-69.3269418280847)(43.0825819210524,-69.4058620794494)(43.1825819210524,-69.484782330814)(43.2825819210524,-69.5637025821787)(43.3825819210524,-69.6426228335434)(43.4825819210524,-69.721543084908)(43.5825819210524,-69.8004633362727)(43.6825819210524,-69.8793835876373)(43.7825819210524,-69.958303839002)(43.8825819210524,-70.0372240903666)(43.9825819210524,-70.1161443417313)(44.0825819210524,-70.1950645930959)(44.1825819210524,-70.2739848444606)(44.2825819210524,-70.3529050958253)(44.3825819210524,-70.4318253471899)(44.4825819210524,-70.5107455985546)(44.5825819210524,-70.5896658499192)(44.6825819210524,-70.6685861012839)(44.7825819210524,-70.7475063526485)(44.8825819210524,-70.8264266040132)(44.9825819210524,-70.9053468553778)(45.0825819210524,-70.9842671067425)(45.1825819210524,-71.0631873581072)(45.2825819210524,-71.1421076094718)(45.3825819210524,-71.2210278608365)(45.4825819210524,-71.2999481122011)(45.5825819210524,-71.3788683635658)(45.6825819210524,-71.4577886149304)(45.7825819210524,-71.5367088662951)(45.8825819210524,-71.6156291176598)(45.9825819210524,-71.6945493690244)(46.0825819210524,-71.7734696203891)(46.1825819210524,-71.8523898717537)(46.2825819210524,-71.9313101231184)(46.3825819210524,-72.010230374483)(46.4825819210524,-72.0891506258477)(46.5825819210524,-72.1680708772123)(46.6825819210524,-72.246991128577)(46.7825819210524,-72.3259113799417)(46.8825819210524,-72.4048316313063)(46.9825819210524,-72.483751882671)(47.0825819210524,-72.5626721340356)(47.1825819210524,-72.6415923854003)(47.2825819210524,-72.7205126367649)(47.3825819210524,-72.7994328881296)(47.4825819210524,-72.8783531394942)(47.5825819210524,-72.9572733908589)(47.6825819210524,-73.0361936422236)(47.7825819210524,-73.1151138935882)(47.8825819210524,-73.1940341449529)(47.9825819210524,-73.2729543963175)(48.0825819210524,-73.3518746476822)(48.1825819210524,-73.4307948990468)(48.2825819210524,-73.5097151504115)(48.3825819210524,-73.5886354017761)(48.4825819210524,-73.6675556531408)(48.5825819210524,-73.7464759045055)(48.6825819210524,-73.8253961558701)(48.7825819210524,-73.9043164072348)(48.8825819210524,-73.9832366585994)(48.9825819210524,-74.0621569099641)(49.0825819210524,-74.1410771613287)(49.1825819210524,-74.2199974126934)(49.2825819210524,-74.2989176640581)(49.3825819210524,-74.3778379154227)(49.4825819210524,-74.4567581667874)(49.5825819210524,-74.535678418152)(49.6825819210524,-74.6145986695167)(49.7825819210524,-74.6935189208813)(49.8825819210524,-74.772439172246)(49.9825819210524,-74.8513594236106)(50.0825819210524,-74.9302796749753)(50.1825819210524,-75.00919992634)(50.2825819210524,-75.0881201777046)(50.3825819210524,-75.1670404290693)(50.4825819210524,-75.2459606804339)(50.5825819210524,-75.3248809317986)(50.6825819210524,-75.4038011831632)(50.7825819210524,-75.4827214345279)(50.8825819210524,-75.5616416858925)(50.9825819210524,-75.6405619372572)(51.0825819210524,-75.7194821886219)(51.1825819210524,-75.7984024399865)(51.2825819210524,-75.8773226913512)(51.3825819210524,-75.9562429427158)(51.4825819210524,-76.0351631940805)(51.5825819210524,-76.1140834454451)(51.6825819210524,-76.1930036968098)(51.7825819210524,-76.2719239481744)(51.8825819210524,-76.3508441995391)(51.9825819210524,-76.4297644509037)(52.0825819210524,-76.5086847022684)(52.1825819210524,-76.5876049536331)(52.2825819210524,-76.6665252049977)(52.3825819210524,-76.7454454563624)(52.4825819210524,-76.824365707727)(52.5825819210524,-76.9032859590917)(52.6825819210524,-76.9822062104564) 
};

\addlegendentry{Linear model (\ref{eq:lineardistance_path_loss})};

\end{axis}
\end{tikzpicture}

%% file: fig5.pdf_tex
\begingroup%
  \makeatletter%
  \providecommand\color[2][]{%
    \errmessage{(Inkscape) Color is used for the text in Inkscape, but the package 'color.sty' is not loaded}%
    \renewcommand\color[2][]{}%
  }%
  \providecommand\transparent[1]{%
    \errmessage{(Inkscape) Transparency is used (non-zero) for the text in Inkscape, but the package 'transparent.sty' is not loaded}%
    \renewcommand\transparent[1]{}%
  }%
  \providecommand\rotatebox[2]{#2}%
  \ifx\svgwidth\undefined%
    \setlength{\unitlength}{420.18022461bp}%
    \ifx\svgscale\undefined%
      \relax%
    \else%
      \setlength{\unitlength}{\unitlength * \real{\svgscale}}%
    \fi%
  \else%
    \setlength{\unitlength}{\svgwidth}%
  \fi%
  \global\let\svgwidth\undefined%
  \global\let\svgscale\undefined%
  \makeatother%
  \begin{picture}(1,0.84513138)%
    \put(0,0){\includegraphics[width=\unitlength]{fig5.pdf}}%
    \put(0.18272807,0.67759899){\color[rgb]{0,0,0}\makebox(0,0)[lb]{\smash{$\mathbf{p}_1$}}}%
    \put(0.24633635,0.53233897){\color[rgb]{0,0,0}\makebox(0,0)[lb]{\smash{$\mathbf{p}_2$}}}%
    \put(0.27937317,0.42526569){\color[rgb]{0,0,0}\makebox(0,0)[lb]{\smash{$\mathbf{p}_3$}}}%
    \put(0.16771443,0.37913808){\color[rgb]{0,0,0}\makebox(0,0)[lb]{\smash{$\mathbf{p}_4$}}}%
    \put(0.29291958,0.26806897){\color[rgb]{0,0,0}\makebox(0,0)[lb]{\smash{$\mathbf{p}_5$}}}%
    \put(0.23056779,0.18729146){\color[rgb]{0,0,0}\makebox(0,0)[lb]{\smash{$\mathbf{p}_6$}}}%
    \put(0.33894431,0.14219066){\color[rgb]{0,0,0}\makebox(0,0)[lb]{\smash{$\mathbf{p}_7$}}}%
    \put(0.42510706,0.17115569){\color[rgb]{0,0,0}\makebox(0,0)[lb]{\smash{$\mathbf{p}_8$}}}%
    \put(0.54761965,0.1704628){\color[rgb]{0,0,0}\makebox(0,0)[lb]{\smash{$\mathbf{p}_9$}}}%
    \put(0.71927189,0.18998405){\color[rgb]{0,0,0}\makebox(0,0)[lb]{\smash{$\mathbf{p}_{10}$}}}%
    \put(0.81822444,0.10718702){\color[rgb]{0,0,0}\makebox(0,0)[lb]{\smash{$\mathbf{p}_{11}$}}}%
    \put(0.90598417,0.06477881){\color[rgb]{0,0,0}\makebox(0,0)[lb]{\smash{$\mathbf{p}_{12}$}}}%
    \put(0.89025106,0.1704628){\color[rgb]{0,0,0}\makebox(0,0)[lb]{\smash{$\mathbf{p}_{13}$}}}%
    \put(0.81889758,0.25191345){\color[rgb]{0,0,0}\makebox(0,0)[lb]{\smash{$\mathbf{p}_{14}$}}}%
    \put(0.70109697,0.33538361){\color[rgb]{0,0,0}\makebox(0,0)[lb]{\smash{$\mathbf{p}_{15}$}}}%
    \put(0.78927905,0.40673709){\color[rgb]{0,0,0}\makebox(0,0)[lb]{\smash{$\mathbf{p}_{16}$}}}%
    \put(0.79264483,0.57502364){\color[rgb]{0,0,0}\makebox(0,0)[lb]{\smash{$\mathbf{p}_{17}$}}}%
    \put(0.89159744,0.64433789){\color[rgb]{0,0,0}\makebox(0,0)[lb]{\smash{$\mathbf{p}_{18}$}}}%
    \put(0.91517724,0.72736588){\color[rgb]{0,0,0}\makebox(0,0)[lb]{\smash{$\mathbf{p}_{19}$}}}%
    \put(0.74889036,0.75092599){\color[rgb]{0,0,0}\makebox(0,0)[lb]{\smash{$\mathbf{p}_{20}$}}}%
    \put(0.02669269,0.5161708){\color[rgb]{0,0,0}\makebox(0,0)[lb]{\smash{$x$}}}%
    \put(0.32915423,0.81492841){\color[rgb]{0,0,0}\makebox(0,0)[lb]{\smash{$y$}}}%
    \put(0.02845319,0.04326088){\color[rgb]{0,0,0}\rotatebox{90}{\makebox(0,0)[lb]{\smash{$5\meter$}}}}%
  \end{picture}%
\endgroup%

%% file: fig6.pdf_tex
\begingroup%
  \makeatletter%
  \providecommand\color[2][]{%
    \errmessage{(Inkscape) Color is used for the text in Inkscape, but the package 'color.sty' is not loaded}%
    \renewcommand\color[2][]{}%
  }%
  \providecommand\transparent[1]{%
    \errmessage{(Inkscape) Transparency is used (non-zero) for the text in Inkscape, but the package 'transparent.sty' is not loaded}%
    \renewcommand\transparent[1]{}%
  }%
  \providecommand\rotatebox[2]{#2}%
  \ifx\svgwidth\undefined%
    \setlength{\unitlength}{400.00002441bp}%
    \ifx\svgscale\undefined%
      \relax%
    \else%
      \setlength{\unitlength}{\unitlength * \real{\svgscale}}%
    \fi%
  \else%
    \setlength{\unitlength}{\svgwidth}%
  \fi%
  \global\let\svgwidth\undefined%
  \global\let\svgscale\undefined%
  \makeatother%
  \begin{picture}(1,0.27224699)%
    \put(0,0){\includegraphics[width=\unitlength]{fig6.pdf}}%
    \put(0.14213652,0.178005){\color[rgb]{0,0,0}\makebox(0,0)[b]{\smash{\small{ML Estimation}}}}%
    \put(0.03628999,0.06420266){\color[rgb]{0,0,0}\makebox(0,0)[lb]{\smash{$\mathsmaller{\left\{ z_{(m,i)}\right\}}$ \footnotesize{(measurements)}}}}%
    \put(0.86610181,0.17800495){\color[rgb]{0,0,0}\makebox(0,0)[b]{\smash{\small{Regression}}}}%
    \put(0.28768716,0.23386896){\color[rgb]{0,0,0}\makebox(0,0)[lb]{\smash{$\mathsmaller{\left\{ \hat{\mu}_{b}(t),\hat{\sigma}_{b}(t)\right\} _{t=1}^{N_{o,\mymax}}}$ \footnotesize{(bias)}}}}%
    \put(0.62492796,0.06258463){\color[rgb]{0,0,0}\makebox(0,0)[lb]{\smash{$\mathsmaller{\left\{ \mu_{b,i}(\mathbf{p}),\sigma_{b,i}^{2}(\mathbf{p})\right\}}$ \footnotesize{(bias)}}}}%
    \put(0.31968731,0.16669119){\color[rgb]{0,0,0}\makebox(0,0)[lb]{\smash{$\mathsmaller{\left\{ \hat{\sigma}_{n}(t)\right\} _{t=1}^{N_{d,\mymax}}}$ \footnotesize{(noise)}}}}%
    \put(0.03622736,0.00593988){\color[rgb]{0,0,0}\makebox(0,0)[lb]{\smash{$\mathsmaller{f(\mathbf{p})}$ \footnotesize{(map)}}}}%
    \put(0.72499811,0.00593897){\color[rgb]{0,0,0}\makebox(0,0)[lb]{\smash{$\mathsmaller{\left\{ \sigma_{n,i}^{2}(\mathbf{p}) \right\}}$ \footnotesize{(noise)}}}}%
    \put(0.14027232,0.22828238){\color[rgb]{0,0,0}\makebox(0,0)[b]{\smash{\small{Step 1:}}}}%
    \put(0.86604517,0.22828238){\color[rgb]{0,0,0}\makebox(0,0)[b]{\smash{\small{Step 2:}}}}%
  \end{picture}%
\endgroup%

%% file: fig9.tikz
%
%
%
\begin{tikzpicture}

\begin{axis}[%
width=\figurewidth,
height=\figureheight,
scale only axis,
xmin=1,
xmax=16,
xlabel={Measurement site index},
ymin=2,
ymax=16,
ytick={2,4,6,8,10,12,14,16},
minor y tick num=4,   
ylabel={RMSE [m]},
legend style={at={(0.03,1)},anchor=north west,nodes=right,draw=none,fill=none,font=\footnotesize,legend columns=2,transpose  legend}]


\addplot [
color=black,
solid,
mark=o,
mark options={solid}
]
table[row sep=crcr]{
1 6.23716476386132\\
2 9.04146038828187\\
3 12.8175565894385\\
4 7.88371231428628\\
5 8.94979665632985\\
6 11.0162222543324\\
7 11.503452837963\\
8 10.4849020461546\\
9 11.3414877177556\\
10 8.92680572365686\\
11 13.0322859062677\\
12 7.56448056449526\\
13 7.53532718360462\\
14 9.17022507495356\\
15 11.1068633185042\\
16 12.7576601035581\\
};
\addlegendentry{$\epsilon_{\ml}^{\rss}$ with $P_{e}^{\nlos}=0.3\qquad$};

\addplot [
color=black,
solid,
mark=asterisk,
mark options={solid}
]
table[row sep=crcr]{
1 6.71209206841001\\
2 8.64092266425076\\
3 10.3515444619896\\
4 5.89287165219379\\
5 4.8807166196328\\
6 7.09192023312945\\
7 9.7014735957577\\
8 6.80037519868678\\
9 10.0009950699083\\
10 6.04640835990614\\
11 7.96539697706003\\
12 6.85926259795293\\
13 6.54609794423794\\
14 6.91730956372446\\
15 10.4451333171567\\
16 8.03191252369361\\
};

\addlegendentry{$\epsilon_{\ml}^{\rss}$ with $P_{e}^{\nlos}=0.1\qquad$};


\addplot [
color=black,
dashed,
mark=diamond,
mark options={solid}
]
table[row sep=crcr]{
1 6.01933333834197\\
2 4.42996959188399\\
3 3.9207419030479\\
4 4.00645376412464\\
5 3.50664601359193\\
6 3.15116863307142\\
7 4.86600312824916\\
8 4.60871677923505\\
9 4.53905490965534\\
10 4.62575278264945\\
11 6.74117120073867\\
12 5.28943666621927\\
13 4.88471719642832\\
14 3.87736140411377\\
15 4.57883245565172\\
16 3.70656194999899\\
};

\addlegendentry{$\epsilon_{\map}^{\rss}$};

\end{axis}
\end{tikzpicture}%

%% file: fig10.tikz
%
%
%
\begin{tikzpicture}

\begin{axis}[%
view={0}{90},
width=\figurewidth,
height=\figureheight,
scale only axis,
xmin=1, xmax=20,
xlabel={Measurement site index},
ymin=0, ymax=6,
xtick={2,4,6,8,10,12,14,16,18,20},
ytick={0,1,2,3,4,5,6},
minor y tick num=4,   
ylabel={RMSE [m]},
legend style={at={(0.97,0.97)},anchor=north east,nodes=right,draw=none,font=\footnotesize, legend columns=2}]


\addplot [
color=black,
dashed,
mark=diamond,
mark options={solid}
]
table[row sep=crcr]{
1 0.399618802408733\\
2 0.226591165160209\\
3 4.15692197419544\\
4 0.174635425590594\\
5 0.302558398492444\\
6 0.242786298966567\\
7 0.155928568186805\\
8 1.35314526295668\\
9 0.530950359739754\\
10 0.388750358949762\\
11 0.318474977712278\\
12 0.0970240010934052\\
13 0.254060286725607\\
14 0.515844950284256\\
15 0.384657236812789\\
16 0.400828319410725\\
17 0.384119959485995\\
18 0.918019283724838\\
19 0.675546402826219\\
20 0.44036064817866\\
};

\addlegendentry{$\epsilon_{\map}^{\toa}$};


\addplot [
color=black,
solid,
mark=asterisk,
mark options={solid}
]
table[row sep=crcr]{
1 0.241347336889444\\
2 0.291247119848477\\
3 3.19505805193503\\
4 0.122953199974606\\
5 0.339626626245288\\
6 0.242523490442203\\
7 0.201169443792595\\
8 1.63871492533379\\
9 0.43873320423469\\
10 0.276188187760813\\
11 0.561161129911269\\
12 1.13999082225494\\
13 0.324358127017315\\
14 0.61755068810189\\
15 0.394650732409079\\
16 0.459492628087398\\
17 0.0431206427035683\\
18 0.923290314210974\\
19 0.451759055779168\\
20 0.188424401290492\\
};

\addlegendentry{$\epsilon_{\ml}^{\toa}$};

\end{axis}
\end{tikzpicture}

%% file: fig11.tikz
%
%
%
%
\begin{tikzpicture}

\begin{axis}[%
width=\figurewidth,
height=\figureheight,
scale only axis,
xmin=0, xmax=8.5,
xlabel={Number of walls $t$},
ymin=0, ymax=40,
ylabel={Bias mean [m]},
axis on top,
legend style={at={(0.03,0.97)},anchor=north west,nodes=right,draw=none,legend columns=1,font=\footnotesize}]

\addplot [
color=black,
only marks,
mark=+,
mark options={solid}
]
plot [error bars/.cd, y dir = both, y explicit]
coordinates{
 (0,0)+-(0.0,0.0)(1,13.255668924342)+-(0.0,2.36561470360943)(2,19.9429940454622)+-(0.0,1.04087046958815)(3,18.7384329508605)+-(0.0,1.23913151141447)(4,19.339218993813)+-(0.0,1.62636010873149)(5,27.014166680741)+-(0.0,2.00167397997721)(6,24.8345429993304)+-(0.0,2.89130685996709)(7,28.9130685996709)+-(0.0,3.71739453424339)(8,28.2218583271289)+-(0.0,8.67392057990126) 
};
\addlegendentry{$\hat{\mu}_{b}(t)$};

\addplot [
color=black,
solid,
line width=1.5pt
]
coordinates{
 (0,0)(1,15.5998249402784)(2,17.6765241989409)(3,19.7532234576034)(4,21.8299227162659)(5,23.9066219749285)(6,25.983321233591)(7,28.0600204922535)(8,30.1367197509161) 
};
\addlegendentry{Model (\ref{eq:bias_mean_RSS})};

\end{axis}
\end{tikzpicture}